\RequirePackage{fix-cm}
\documentclass[twocolumn]{svjour3}          
\smartqed  
\usepackage{graphicx}
\usepackage{subfigure}
\usepackage{mathptmx}      
\usepackage{epstopdf}
\usepackage{cite}
\usepackage{amsmath}
\usepackage{mathrsfs}
\usepackage{amssymb}
\usepackage{bm}
\usepackage[noend]{algpseudocode}
\usepackage{algorithmicx,algorithm}
\usepackage{threeparttable}
\usepackage{booktabs}
\usepackage{marvosym}
\usepackage{balance}
\usepackage{CJKutf8}   

\usepackage[T1]{fontenc}
\usepackage[utf8]{inputenc}

\usepackage{newtxtext,newtxmath}

\usepackage{tikz,xcolor}
\usepackage{hyperref}
\hypersetup{
	colorlinks=true,
	linkcolor=blue,
	filecolor=blue,      
	urlcolor=blue,
	citecolor=blue,
}

\definecolor{lime}{HTML}{A6CE39}
\DeclareRobustCommand{\orcidicon}{
\begin{tikzpicture}
\draw[lime, fill=lime] (0,0)
circle[radius=0.16]
node[white]{{\fontfamily{qag}\selectfont \tiny \.{I}D}}; 
\end{tikzpicture}
\hspace{-2mm}
}
\foreach \x in {A, ..., Z}{%
\expandafter\xdef\csname orcid\x\endcsname{\noexpand\href{https://orcid.org/\csname orcidauthor\x\endcsname}{\noexpand\orcidicon}}
}

\spnewtheorem{assumption}{Assumption}{\bf}{\rm}
\graphicspath{{Figures/}}
\journalname{Journal}

\begin{document}
\begin{CJK}{UTF8}{gbsn}

\title{Energy-storing analysis and fishtail stiffness optimization \\ for a wire-driven elastic robotic fish}

\author{Xiaocun~Liao\orcidA{} \and
        Chao~Zhou\orcidB{} \and
        Junfeng~Fan\orcidC{} \and
        Zhuoliang~Zhang\orcidD{} \and
       Zhaoran~Yin\orcidE{}  \and
       Liangwei~Deng\orcidF{}   
}
    
\authorrunning{
        X.~Liao \and 
        C.~Zhou \and 
        J.~Fan \and
        Z.~Zhang \and 
        Z.~Yin \and 
        L.~Deng
    }

\institute{X.~Liao \and C.~Zhou(\Letter) \and J.~Fan \and Z.~Zhang \and Z.~Yin \and L.~Deng
\at The Laboratory of Cognition and Decision Intelligence for Complex Systems, Institute of Automation, Chinese Academy of Sciences, Beijing 100190, China \\
\email{chao.zhou@ia.ac.cn}
\and X.~Liao \and Z.~Zhang \and Z.~Yin \and L.~Deng \at
School of Artificial Intelligence, University of Chinese Academy of Sciences, Beijing 100049, China}

\date{Received: date / Accepted: date}

\maketitle

\begin{abstract}
The robotic fish with high propulsion efficiency and good maneuverability achieves underwater fishlike propulsion by commonly adopting the motor to drive the fishtail, causing the significant fluctuations of the motor power due to the uneven swing speed of the fishtail in one swing cycle. Hence, we propose a wire-driven robotic fish with a spring-steel-based active-segment elastic spine. This bionic spine can produce elastic deformation to store energy under the action of the wire driving and motor for responding to the fluctuations of the motor power. Further, we analyze the effects of the energy-storing of the active-segment elastic spine on the smoothness of motor power. Based on the developed Lagrangian dynamic model and cantilever beam model, the power-variance-based nonlinear optimization model for the stiffness of the active-segment elastic spine is established to respond to the sharp fluctuations of motor power during each fishtail swing cycle. Results validate that the energy-storing of the active-segment elastic spine plays a vital role in improving the power fluctuations and maximum frequency of the motor by adjusting its stiffness reasonably, which is beneficial to achieving high propulsion and high speed for robotic fish. Compared with the active-segment rigid spine that is incapable of storing energy, the energy-storing of the active-segment elastic spine is beneficial to increase the maximum frequency of the motor and the average thrust of the fishtail by 0.41 Hz, and 0.06 N, respectively.

\end{abstract}

\section{Introduction}
\label{sec:Introduction}
There are many abundant resources in the ocean for humans to be exploited. With the development of robotics, various robots have been applied to replace humans in underwater work. Compared with the traditional propeller-based underwater robot, the bionic robotic fish features high propulsion efficiency and good maneuverability by imitating fish swimming, and holds the potential prospects in underwater applications \cite{Scaradozzi_2017_OE, Zhang_2020_NODY}.

Recently, various robotic fishes, which are usually based on the discrete joints \cite{Omari_2021_OE, Clapham_2015_Springer}, fluid-driven mode \cite{Liu_2022_BB, Aubin_2019_Nature, Marchese_2014_SoftRobotics, Katzschmann_2018_SciRob}, smart material \cite{Wang_2008_SMS, Li_2022_SMS, Li_2021_Nature, Ning_2022_MRS}, magnetic actuator \cite{Chen_2020_SMC, Huang_2021_BB}, and tensegrity joint \cite{Chen_2019_SciRob, Chen_2021_TRO}, have been developed successfully. Clapham \emph{et al.} proposed the robotic fish named as iSplash-I, which reached a maximum frequency of 6.6 Hz and a maximum speed of 2.8 BL/s. Based on the iSplash-I, the iSplash-II was further developed, which featured a maximum frequency of 20 Hz and a maximum speed of 11.6 BL/s \cite{Clapham_2015_Springer}. Using fluidic elastomer actuators, Marchese \emph{et al.} developed a soft robotic fish capable of escape maneuvers \cite{Marchese_2014_SoftRobotics}. A self-powered soft robot driven by the dielectric elastomer actuator was applied to explore the Mariana Trench \cite{Li_2021_Nature}. In addition, the wire-driven robotic fish has also obtained extensive attention \cite{Zheng_2013_ICIRS, Zhong_2017_Tmech, LiuJ_2021_BB, Shintake_2020_ICRA, Estarki_2021_ICRoM}. Li \emph{et al.} developed a wire-driven robotic fish, whose fishtail could swing in any direction \cite{Zheng_2013_ICIRS}. The fishtail consisted of seven vertebraes that were connected by ball hinges, and two pairs of wires were perpendicularly mounted to drive the fishtail. A wire-driven robotic fish with an active body and compliant tail was proposed by Zhong \emph{et al.}, and it was capable of an average turning speed of 63 $^{\circ}$/s and a maximum swimming speed of 2.15 BL/s \cite{Zhong_2017_Tmech}. Shintake \emph{et al.} presented a wire-driven robotic fish with tensegrity systems, which could tune body stiffness by adjusting the cross-section and prestretch ratio of the cables \cite{Shintake_2020_ICRA}. 

To improve the swimming performance of robotic fish, such as swimming speed and propulsion efficiency, many efforts have focused on the exploration of the passive-segment elastic spine (PES) \cite{Park_2010_ICBRB, Reddy_2018_MMT, Chen_2022_Tmech}. With the interaction of fishtail and fluid, the PES can store and release the energy periodically owing to the elastic deformation, which is beneficial to the improvement of propulsion performance. Hence, the stiffness of the PES can be tuned to improve the propulsion performance of robotic fish. Park \emph{et al.} explored the influences of the PES on the thrust, and verified that there was an optimal stiffness that could maximize the thrust at the given frequency \cite{Park_2010_ICBRB}. Inspired by the caudal fin of fish, Reddy N \emph{et al.} studied the influence of the stiffness distribution of the passive caudal fin on the thrust, and determined the optimal stiffness distribution of the caudal fin based on the developed optimization model \cite{Reddy_2018_MMT}. Chen \emph{et al.} adopted double torsion springs to design the PES, and verified that the PES was beneficial to obtain high swimming speed, high thrust, and low cost of transport (COT) \cite{Chen_2022_Tmech}. 

To imitate the continuum fishtail, the active-segment elastic spine (AES) has also been widely used in robotic fish \cite{LiZ_2013_AMM, Lau_2015_RAM, ElDaou_2012_ICRA, Valdivia_2006_JDS, Fujiwara_2017_Abmech, Jiang_2019_CAC}. Different from the PES, the elastic deformation of the AES comes into being owing to the active action of the drive mechanism, and the deformation magnitude is controllable. The drive mechanisms including wire driving, bar driving, etc., are usually powered by the motor. For example, Li \emph{et al.} designed a wire-driven robotic fish with a continuum fish tail, whose fishlike spine was an elastic-plate-based active-segment elastic spine \cite{LiZ_2013_AMM}. The robotic fish was driven by a wire pair to obtain the C-shape swing, and was capable of 0.254 BL/s. A robotic shark that adopted an elastic beam and plate to simulate the fish spine is developed by Lau \emph{et al.}, and the elastic beam serving as the AES was driven by two pairs of wires \cite{Lau_2015_RAM}. Due to the double-wire driving, the robotic shark, whose fishtail could swing asymmetrically, was capable of achieving forward motion as well as ascending motion. To imitate carangiform swimming, a robotic fish, whose continuum fishtail contained a section of the polystyrene-chloride-sheet-based active-segment elastic spine and was driven by the L-shape metal bar and wire, was designed by Fujiwara \emph{et al} \cite{Fujiwara_2017_Abmech}. Some similar researches can be found in the literatures \cite{ElDaou_2012_ICRA, Valdivia_2006_JDS, Jiang_2019_CAC}. Although the AES has been widely applied for pursuing continuum fishtail in robotic fish, the researches involving the detailed analysis of the new advantages brought by the AES, and the optimization of the AES are worthy of further exploration.

For robotic fish, the motor power features the periodic sinusoid-like law due to the characteristics of the biomimetic undulation motion. Hence, the motor power undulates sharply under external disturbance, and the instantaneous maximum power exceeds the allowable power value of the motor easily, which causes damage to the motor. Moreover, the motor performance is also limited, which hinders the improvement of the maximum swing frequency and swimming speed of robotic fish.

The main purpose of this paper is to solve the fluctuations of the motor power by optimizing the stiffness of the AES to cut down the power variance. The AES equipped by the robotic fish can dent the peak value and fill the valley value for motor power by energy storage and release in one swing cycle, which can respond to the fluctuations of the motor power effectively, and protect the motor from damage. But, how to maximize the advantages of the AES's energy-storing is challenging. In this paper, we focus on a wire-driven elastic robotic fish, whose fishlike tail is based on dual spring steel and includes a AES and a PES. We analyze the effects of the AES's energy-storing on the smoothness of motor power. Based on the Lagrangian dynamic model and the cantilever beam model, a power-variance-based nonlinear optimization model of the AES's stiffness is established to pursue stable motor power, where the swing frequency and PES's stiffness are taken into account. Extensive simulations are conducted, and validate that the AES's energy-storing and stiffness have no contribution to the average power of the motor, but affect the smoothness of motor power. Compared with the active-segment rigid spine (ARS) that is incapable of storing energy, the AES's stiffness can be tuned to make the motor power more stable owing to the periodic energy-storing, which is beneficial to the improvement of the maximum frequency of the motor and the swimming performance of robotic fish.The contributions of this paper are as follows.
\begin{enumerate}
\item A dynamic model without the consideration of the periodic vibration of the fish head is proposed, which provides an efficient tool for energy-storing analysis and stiffness optimization of AES.
\item The effects of AES's energy-storing on motor power in terms of mean and variance are analyzed, and the AES's stiffness is further optimized to pursue more smooth and stable motor power.
\end{enumerate}

The rest of this paper is organized as follows. The mechanism design and dynamic modeling of robotic fish are provided in section \ref{sec:Mechatronic}. Section \ref{sec:StiffnessOptimization} presents the analysis of energy storage and the nonlinear stiffness optimization for the AES. The results are offered in section \ref{sec:Simulation}. Finally, section \ref{sec:Conclusions} summarizes the conclusions and future works.

\section{Mechanism design and dynamic modeling}
\label{sec:Mechatronic}
This section introduces the mechanism design and dynamic model of the wire-driven elastic robotic fish, which lays a foundation for the following optimization.

\subsection{Design of robotic fishtail}
\label{Sec:MechatronicDesign}

\begin{figure}[t]
    \centering
    \includegraphics[width=7.5cm]{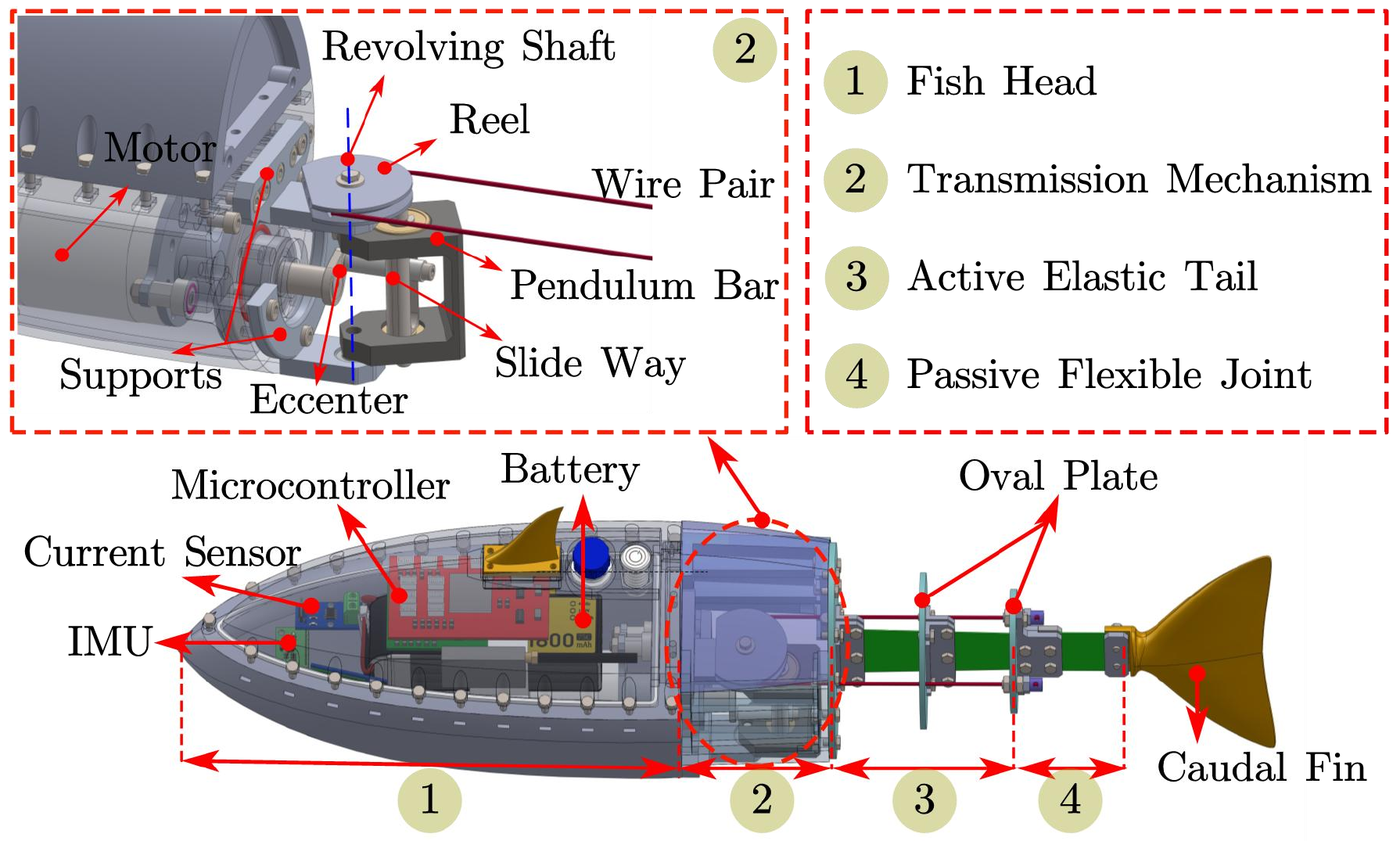}
    \caption{Overview design of the wire-driven elastic robotic fish.}
    \label{fig:fishMechanism}
\end{figure}

To emulate the fish swimming, a wire-driven robotic fish has been developed in our previous work \cite{Liao_2023_Tmech}, which includes a rigid fish head, an efficient transmission mechanism, an active elastic tail, a passive flexible joint, and a rigid caudal fin, as shown in Fig. \ref{fig:fishMechanism}. The transmission mechanism plays a significant role in transforming the circumferential rotation of the motor into the fishlike back-and-forth swing. Concretely, the L-shaped eccenter, whose one end is inserted into the slide way and other end is connected to the output shaft of the motor, is driven by the motor. The eccenter further drives the pendulum bar by the slide way to obtain the fishlike swing, and the reel also rotates along with the pendulum bar owing to the fixed connection. 

The active elastic tail and passive flexible joint are based on the active-segment elastic spine and passive-segment elastic spine, respectively, which are made of spring steel with energy-storing. The AES is based on a trapezoidal spring steel and driven by the wire. The PES is a rectangular spring steel. When the reel rotates back-and-forth, the wire's length on both sides of the fishtail changes periodically, resulting in the fishlike swing of the elastic fishtail. When the fishtail swings, the PES periodically produces elastic deformations to store energy due to the action of hydrodynamic force, which is beneficial to the improvement of swimming performance, such as swimming speed and propulsion efficiency. Hence, the PES's stiffness can be tuned to affect its elastic deformation and the swimming performance of robotic fish. 

Differing from the PES, the AES can generate elastic deformation to achieve periodic energy-storing under the action of the wire. Assuming that the output torque of the motor and the tension of the wire are large enough, the magnitude of the elastic deformation for the AES does not depend on its stiffness. That is because the deformation of AES depends on the change law of the wire length. As long as the length of the wire on both sides of the fishtail is given, the deformation of AES can be uniquely determined, regardless of its stiffness. It means that the swing law of the fishtail is also independent of the AES's stiffness. In other words, the AES's stiffness does not affect the swimming speed of the robotic fish. Given that the elastic potential energy of the AES is provided by the motor, this paper focuses on the exploration of the advantages of the AES's energy-storing on the motor power of robotic fish, and realizes the optimization design of AES's stiffness. 

\subsection{Dynamic model}


For the bionic swing of the robotic fish, the periodic vibration of the fish head plays a buffering role, which results in the decreases in the motor output torque and power, to a certain extent. To fully ensure the reliability and stability of the robotic fish, the periodic vibration of the fish head needs to be ignored and the maximum load of the motor needs to be considered. Further, the most reliable solution for stiffness adjustment is provided by analyzing the maximum motor power. Based on the dynamic model of untethered swimming presented in previous work \cite{Liao_2023_Tmech}, a dynamic model without the consideration of the periodic vibration of the fish head is proposed in this section.

\subsubsection{Coordinate frames}

In response to the continuum fishtail, we make two assumptions as follows. Firstly, the shape of AES is an arc after bending, which can be replaced by the corresponding chord in the dynamic model. Secondly, the PES is modeled by a torsion spring since its length is far less than the body length.

\begin{figure}[t]
    \centering
    \includegraphics[width=8cm]{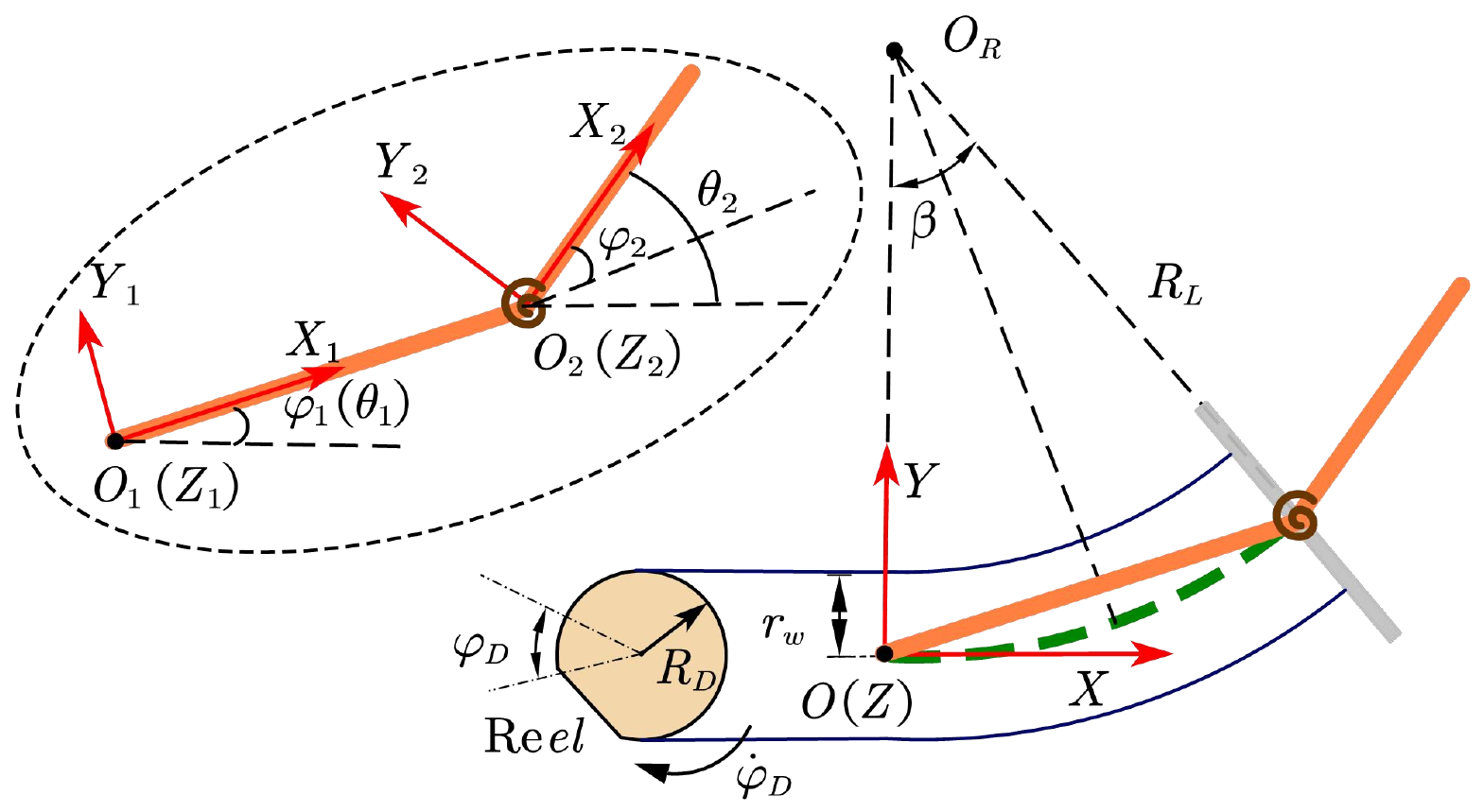}
    \caption{Definition of coordinate frames. The green dotted curve denotes the AES. The two orange solid lines represent the chord of the AES and the caudal fin, respectively. The PES is simplified as the torsion spring.}
    \label{fig:Dynamic}
\end{figure}

The inertial reference frame $C_w=\{ O-XYZ \}$ and link-fixed reference frame $C_i=\{ O_i-X_iY_iZ_i \}$ ($i$ = 1, 2) that is attached to link $L_i$ are defined in the Fig. \ref{fig:Dynamic}. Note that the AES is denoted as $\hat{L}_1$, and the $L_1$ and $L_2$ denote the chord of AES and the caudal fin, respectively. $l_i$, $w_i$, $h_i$ and $m_i$ represent the length, width, height and mass of the link $L_i$, respectively. $l_{T,i}$, $w_{T,i}$, and $d_{T,i}$ denote the length, width, thickness of the \emph{i}-th spring steel ($i=1$: AES, $i=2$: PES), respectively. $E_{i}$, and $I_{Zi}$ denote the elastic modulus and the area moment of inertia of the \emph{i}-th spring steel, respectively. The angle of the axis $X$ and $X_i$ is denoted as $\theta_i$. $\varphi_1$ is the angle between the axis $X$ and $X_1$, and $\varphi_2$ is the angle between the axis $X_2$ and $X_1$. $\beta$ denotes the central angle of $\hat{L}_1$ and is positive when AES bends to the right.

\subsubsection{Kinematics model}

According to the transmission mechanism, the angle of the reel is calculated by
\begin{equation}
\label{eq:varphiD}
\varphi_D = \arcsin \left( d_1 \sin (\varphi_m)/d_2 \right), 
\end{equation}
where $d_1$ and $d_2$ represent the bias of the eccenter and the distance between the revolving shaft and slide way, respectively;  $\varphi_m$ denotes the angle of motor and is presented by $\varphi_m = \omega t$; $\omega$ is the angular velocity of the motor. Based on the Fig. \ref{fig:Dynamic}, we can obtain
\begin{equation}
\label{eq:beta}
    \beta = R_D\varphi_D/r_w,
\end{equation}
\begin{equation}
    l_1 = 2R_L \sin \left( \beta/2 \right) = 2 l_{T,1} \sin \left( \beta/2 \right) /\beta,
\end{equation}
where $R_D$ and $r_w$ denote the radius of the reel and the distance between the body axis and wire pair, respectively; $R_L$ is the arc radius of the AES. The $\theta_1$ and $\theta_2$ are presented by $\theta_1 = \varphi_1 = \beta/2$, $\theta_2 = \theta_{1} + \varphi_2$, respectively.

Further, we define the generalized vector as $\boldsymbol{q} = \left[\varphi_1 \ \varphi_2 \right] ^T$, and formulate the velocity and angular velocity of the center of mass of the link $L_i$ as follows:
\begin{equation}
    ^w\boldsymbol{v}_1 = \boldsymbol{J}_{v1}\boldsymbol{\dot q},\ ^w\boldsymbol{v}_2 = \boldsymbol{J}_{v2}\boldsymbol{\dot q},
\end{equation}
\begin{equation}
    ^w\boldsymbol{\omega}_1 = \boldsymbol{J}_{\omega 1}\boldsymbol{\dot q},\ ^w\boldsymbol{\omega}_2 = \boldsymbol{J}_{\omega 2}\boldsymbol{\dot q},
\end{equation}
\begin{equation}
    \boldsymbol{J}_{v1} = 
    \begin{aligned}
        \left[ 
                \begin{array}{cc}
                - l_{c,1} \sin q_1& 0 \\
                l_{c,1} \cos q_1 & 0 \\
                0 & 0 
            \end{array} 
        \right]
\end{aligned},
\end{equation}
\begin{equation}
\begin{aligned}
     \boldsymbol{J}_{v2} =  \begin{aligned}
        \left[ 
                \begin{array}{cc}
                -l_1 \sin q_1 - l_{c,2} \sin (q_1+q_2)& -l_{c,2} \sin (q_1+q_2) \\
                l_1 \cos q_1 + l_{c,2} \cos (q_1+q_2) & l_{c,2} \cos (q_1+q_2) \\
                0 & 0 
            \end{array} 
        \right]
\end{aligned},
\end{aligned}
\end{equation}
\begin{equation}
    \boldsymbol{J}_{\omega 1} = 
    \begin{aligned}
        \left[ 
                \begin{array}{cc}
                0 & 0 \\
                0 & 0 \\
                1 & 0 
            \end{array} 
        \right]
        \end{aligned},\ 
        \boldsymbol{J}_{\omega 2} = 
    \begin{aligned}
        \left[ 
                \begin{array}{cc}
                0 & 0 \\
                0 & 0 \\
                1 & 1 
            \end{array} 
        \right]
\end{aligned},
\end{equation}
where $\boldsymbol{J}_{vi}$ and $\boldsymbol{J}_{\omega i}$ are the Jacobi matrix of the link $L_i$; $l_{c,i}$ denotes the distance between the center of mass of the link $L_i$ and $O_i$.

\subsubsection{Lagrangian dynamic modeling}
The Lagrangian function $\Upsilon$ is defined as the difference between the kinetic energy $E_k$ and the potential energy $E_p$, which is calculated by 
\begin{equation}\label{eq:LagrangianFunction}
\Upsilon = E_k - E_p,
\end{equation}
\begin{equation}
E_k = \frac{1}{2} \sum _{i=1}^{2} \left(\ ^w\boldsymbol{v}_i^T \boldsymbol{M}_{i}\ ^w\boldsymbol{v}_i + \ ^w \boldsymbol{\omega}_i^T \boldsymbol{I}_{i}\ ^w \boldsymbol{\omega}_i \right),
\end{equation}
\begin{equation}
\label{eq:E_pes}
E_{p} = E_{pes} = \frac{1}{2} \kappa_{2} \theta_s^2,
\end{equation}
where $\boldsymbol{M}_{i}$ and $\boldsymbol{I}_{i}$ denote the mass matrix and the inertia tensor of $L_i$, respectively; $E_{pes}$ represents the elastic potential energy of the PES; $\theta_s$ represents the PES's bend angle and is equal to $\varphi_2 -\beta/2$; $\kappa_{2}$ denotes the torsional spring constant or the PES's stiffness, and is calculated by
\begin{equation}
\label{eq:K_pes}
\kappa_{2} = \frac{E_2I_{Z2}}{l_{T,2}} = \frac{E_2 w_{T,2} d_{T,2}^3}{12 l_{T,2}}.
\end{equation}
As presented in the above analysis of section \ref{Sec:MechatronicDesign}, the AES's stiffness does not affect the swimming speed of the robotic fish, which is why the Lagrangian function excludes the AES's elastic potential. 

Finally, the dynamic model without the consideration of the periodic vibration of the fish head is equal to 
\begin{equation} \label{eq:dynamicmodel}
\frac{d}{dt} \frac{\partial \Upsilon}{\partial \dot{\boldsymbol{q}}} -\frac{\partial \Upsilon}{\partial \boldsymbol{q}} =[\tau_{J1} + \tau_{Hd,1} \  \tau_{Hd,2} ]^T,
\end{equation}
where $\tau_{J1}$ denotes the equivalent joint torque used to drive the link $L_1$, and $\tau_{Hd,i}$ represents the generalized torque of the hydrodynamic force with respect to $O_i$.

\subsubsection{Hydrodynamic force analysis}

According to the Morison equation, the added mass force $^w\boldsymbol{F}_{a,i}$ and the drag forces $^w\boldsymbol{F}_{d,i}$ of the link $L_i$ are calculated by
\begin{equation}
   ^w\boldsymbol{F}_{a,i} = -m_{a,i}\ ^w\dot{\boldsymbol{v}}_{i},
\end{equation}
\begin{equation}
\renewcommand\arraystretch{1.5}
   ^w\boldsymbol{F}_{d,i} =\ ^w \boldsymbol{R} _i
   \begin{aligned}
        \left[ 
                \begin{array}{c}
                -\frac{1}{2} \rho c_{f,i} S_{x,i}\ ^iv_{x,i} \lvert ^iv_{x,i} \lvert \\
                -\frac{1}{2} \rho c_{d,i} S_{y,i}\ ^iv_{y,i} \lvert ^iv_{y,i} \lvert \\
                0
            \end{array} 
        \right]
        \end{aligned},
\end{equation}
respectively, where the added mass $m_{a,i}$ is equal to the product of $m_i$ and the added mass coefficient $c_{m,i}$; $^w\dot{\boldsymbol{v}}_{i}$ and $^i \boldsymbol{v}_{i} = [^i \boldsymbol{v}_{x,i} \ ^i \boldsymbol{v}_{y,i} \ 0 ]^T$ are the acceleration and velocity of the center of mass of $L_i$, respectively, and the left superscript $w$ and $i$ denote the reference frame $C_w$ and $C_i$, respectively; $S_{x,i}$ and $S_{y,i}$ denote the characteristic areas of $L_i$ on the $X_i$ and $Y_i$, respectively; $c_{d,i}$ and $c_{f,i}$ denote the drag coefficient and friction coefficient of $L_i$, respectively; $\rho$ denotes the fluid density; $^w\boldsymbol{R}_i$ denotes the rotation matrix of $C_i$ with respect to $C_w$, and can be presented by
\begin{equation}
\begin{aligned}
^{w}\boldsymbol{R}_i =\left[ \begin{array}{ccc}
\cos\theta_i & - \sin\theta_i&0\\
\sin\theta_i & \cos\theta_i& 0\\
0 & 0 & 1
\end{array}  \right]
\end{aligned}.
\end{equation}

Hence, the hydrodynamic force of the link $L_i$ is equal to 
\begin{equation}
   ^w\boldsymbol{F}_{ad,i} = \ ^w\boldsymbol{F}_{a,i} + \ ^w\boldsymbol{F}_{d,i}.
\end{equation}

Further, the $\tau_{Hd,1}$ and $\tau_{Hd,2}$ can be expressed as follows:
\begin{equation}
\label{eq:tauHd1}
   \tau_{Hd,1} = \boldsymbol{k}^T (^w\boldsymbol{r}_{cm1} \times \ ^w \boldsymbol{F}_{ad,1} +\ ^w \boldsymbol{r}_{cm2} \times \ ^w\boldsymbol{F}_{ad,2}),
\end{equation}
\begin{equation}
\label{eq:tauHd2}
   \tau_{Hd,2} = \boldsymbol{k}^T \left [ (^w \boldsymbol{r}_{cm2} -\ ^w\boldsymbol{r}_{O_2}) \times \ ^w\boldsymbol{F}_{ad,2} \right],
\end{equation}
where $^w\boldsymbol{r}_{cmi}$ denotes the position of the center of mass of the link $L_i$ in the $C_w$; $^w\boldsymbol{r}_{O_2}$ represents the position of $O_2$ in the $C_w$; $\boldsymbol{k}$ is equal to $[0 \ 0 \ 1]^T$.

Based on the Eq. (\ref{eq:dynamicmodel}), (\ref{eq:tauHd1}) and (\ref{eq:tauHd2}), we can obtain the real-time states of the robotic fish, e.g., $\theta_s$, $\tau_{J1}$, which are affected by the frequency $f$, the PES's stiffness $\kappa_{2}$. Based on the above dynamic model, the output torque and power of the motor can be obtained further by incorporating the force models of the AES and transmission mechanism, and the power-variance-based nonlinear optimization model can be established to optimize the AES's stiffness.

\section{Energy-storing analysis and stiffness optimization model for AES}
\label{sec:StiffnessOptimization}
This section focuses on the functions of the AES's energy-storing, and further derives the nonlinear optimization model of the AES's stiffness by combining the developed dynamic model, the force models of the AES and transmission mechanism. Note that the AES's stiffness is calculated by
\begin{equation}
\label{eq:K_aes}
\kappa_{1} = \frac{E_1I_{Z1}}{l_{T,1}} = \frac{E_1 w_{T,1} d_{T,1}^3}{12 l_{T,1}}.
\end{equation}
We adjust the AES's stiffness by changing its thickness $d_{T,1}$ in this paper. The thickness resolution of the spring steel is 0.1 mm.

\subsection{Analysis for the AES’s energy-storing}

More attractively, the AES can store and release energy periodically due to elastic deformation. Based on the model of the cantilever beam, the bending moment and the elastic potential energy of the AES are equal to
\begin{equation}
    T_{e,1} = \kappa_{1} \beta  = \frac{\beta E_1 w_{T,1} d_{T,1}^3}{12 l_{T,1}},
\end{equation}
\begin{equation}
\label{eq:AESEnergy}
    E_{aes} = \frac{1}{2} \kappa_{1} \beta ^2 =  \frac{\beta^2 E_1 w_{T,1} d_{T,1}^3}{24 l_{T,1}},
\end{equation}
respectively. 

\begin{figure}[t]
    \centering
    \includegraphics[width=7.5cm]{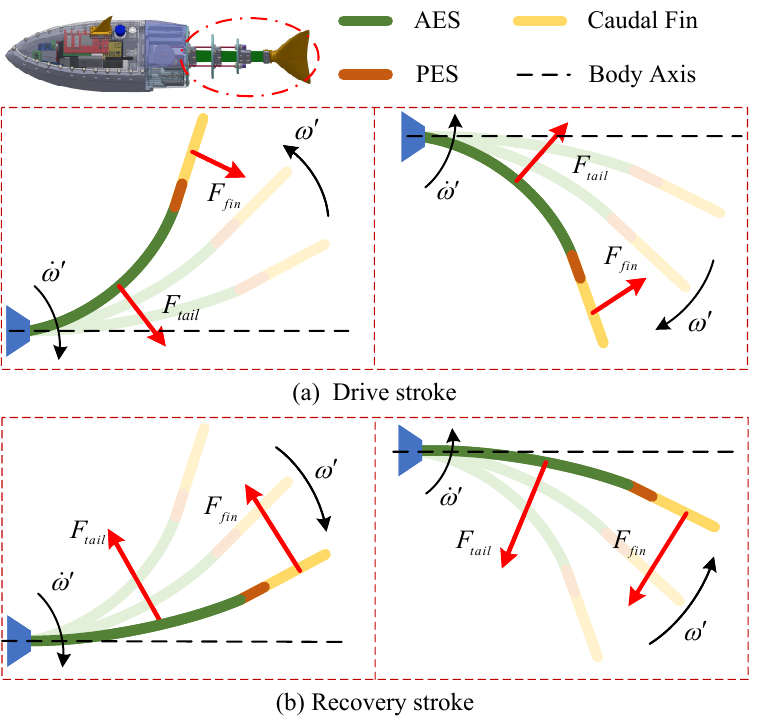}
    \caption{Swing analysis of fishtail in one cycle. The swing cycle of fishtail includes (a) drive stroke and (b) recovery stroke. The drive stroke is defined as the duration when the fishtail swings from the middle to the sides, and the recovery stroke is opposite to the drive stroke. The $F_{tail}$ and $F_{fin}$ represent the hydrodynamic forces of the fishtail and caudal fin, respectively.}
    \label{fig:AnalysisInoneCycle}
\end{figure}

To explore the effects of AES’s energy-storing on the motor power, the swing process of the fishtail in one cycle is analyzed, as shown in Fig. \ref{fig:AnalysisInoneCycle}. It's obvious that the angular acceleration $\dot{\omega}^{\prime}$ points to the initial position of the body axis owing to the sinusoid-like swing of the fishtail. The direction of the angular acceleration and hydrodynamic forces (including the $F_{tail}$ and $F_{fin}$) keep coincident in the drive stroke, and opposite in the recovery stroke, which results in the difference of motor torque in the two strokes. Concretely, the motor torque during the drive stroke is significantly less than that in the recovery stroke, and the motor power also features a similar law to the motor torque.

Interestingly, the output power of the motor increases obviously during the drive stroke since the AES needs to store energy provided by the motor. On the contrary, the AES returns to the equilibrium state and releases the stored elastic potential energy during the recovery stroke, which can reduce the output power of the motor. Hence, for the whole swing cycle, the output power of the motor becomes more stable owing to the AES's energy-storing. Given that the energy-storing capability of the AES is directly related to its stiffness, we establish the stiffness optimization model with the purpose of improving the smoothness of the output power of the motor.

\subsection{Nonlinear optimization model for AES’s stiffness}
Based on the above dynamic model, we can obtain the equivalent joint torque $\tau_{J1}$. Obviously, the equivalent torque generated from the pulling force of the wire is equal to the sum of the equivalent joint torque of the link $L_1$ and the bending moment of the AES, that is,
\begin{equation}
    T_{wire,eq} = \tau_{J1} + T_{e,1}.
\end{equation}

Hence, the pulling force of the wire is equal to 
\begin{equation}
    F_{wire} = \frac{T_{wire,eq}}{r_w},
\end{equation}
where the positive and negative $F_{wire}$ represent the pulling force of the right and left wires, respectively. The torque generated from the wire with respect to the reel is equal to 
\begin{equation}
    T_{wire,r} = F_{wire}R_D.
\end{equation}

Taking the reel, slide way and pendulum bar as a whole object, and ignoring the self-rotation of the slide way, we can formulate the dynamic model of the transmission mechanism according to the law of rotation, that is
\begin{equation}
\label{eq:LawRotation}
    -\frac{T_m}{d_1 \cos \varphi_m}d_2 \cos \varphi_D + T_{wire,r} = (J_R + J_S + J_P)\Ddot{\varphi}_D,
\end{equation}
where $J_R$, $J_S$, and $J_P$ denote the moment of inertia of the reel, slide way, and pendulum bar, respectively; $T_m$ represents the motor torque that is presented by
\begin{equation}
\label{eq:Tm}
    T_m = - \frac{-T_{wire,r} + (J_R + J_S + J_P)\Ddot{\varphi}_D}{d_2 \cos \varphi_D}d_1 \cos \varphi_m.
\end{equation}

To make the motor power more smooth, we establish the following nonlinear optimization model that is to minimize the variance of the motor power, that is 
\begin{equation}
\label{eq:minPowerVarNoLimit}
    \min \limits_{\kappa_{1}} var (P_m(t)) \| _0 ^t,
\end{equation}
where $var (\cdot) \| _0 ^t$ represents the variance of $\cdot$ within the duration of $0-t$, and $P_m = T_m \omega$ denotes the motor power. It is evident that when the average power of motor $\bar P_m$ remains constant, the smaller the variance of the motor power is, the more stable the motor power is, meaning that the peak-to-peak value and peak value of the motor power are smaller. Hence, it is beneficial to protect the motor from damage and improve the maximum output frequency. Similarly, we can adopt the optimization model as follows:
\begin{equation}
\label{eq:minMaxDiffMin}
    \min \limits_{\kappa_{1}} \left\{ \max (P_m(t)) \| _0 ^t - \min (P_m(t)) \| _0 ^t \right\},
\end{equation}
where $\max (\cdot) \| _0 ^t$ and $\min (\cdot) \| _0 ^t$ represent the maximum and minimum values of $\cdot$ within the duration of $0-t$, respectively. Specially, we can determine the AES's stiffness range by 
\begin{equation}
    \kappa_{1} \in \left\{ \kappa_{1} \lvert var (P_m(t)) \| _0 ^t \leq var_{m} \right\},
\end{equation}
where $var_{m}$ denotes the allowed maximum variance of motor power.

\subsection{Determination of the AES’s stiffness range}

For the above optimization model, the reasonable range of $\kappa_{1}$ should be determined. To calculate the maximum $\kappa_{1}$, the statics is adopted, that is $\tau_{J1} = 0$ and $\Ddot{\varphi}_D = 0$. The motor torque in the statics can be expressed by 
\begin{equation}
\label{eq:staticsTorque}
    T_m^s = \frac{T_{e,1} R_D}{r_w d_2 \cos \varphi_D} d_1 \cos \varphi_m.
\end{equation}

Obviously, the maximum $\kappa_{1}$ is determined by 
\begin{equation}
\label{eq:maxdT1}
\begin{aligned}
    & \kappa_{1}^{max}=\max \{ \kappa_{1} \}\\
    & s.t.\  \max(T_{m}^s)\le T_{m}^{max}
\end{aligned},
\end{equation}
where $T_{m}^{max}$ denotes the maximum torque that the motor can provide.

On the other hand, the AES isn't capable of supporting the fishtail owing to the axial force from the caudal fin, if the small $\kappa_{1}$ is used. In this case, there is an unexpected deformation for the AES, which affects the C-shape swing and amplitude of the fishtail. In this paper, we utilize the model of buckling of columns to determine the minimum $\kappa_{1}$ at the given $f$ and $\kappa_{2}$, that is, 
\begin{equation}
\label{eq:mindT1}
\begin{aligned}
    \kappa_{1}^{min}(f,\kappa_{2}) &= \min\{ \kappa_{1} \}\\
    s.t.\ \frac{\pi^2E_1I_{Z1}}{(\mu l_{T,1})^2} & \geq F_{cr}
\end{aligned}.
\end{equation}
where $\mu$ represents the length factor and is equal to 2; $F_{cr}$ denotes the axial force generated from the caudal fin at the free swimming, and is calculated by
\begin{equation}
    F_{cr} = \boldsymbol{i}^{T}[\  {^1 \boldsymbol{R}_w} (^w \boldsymbol{F}_{a,2}\ + \ ^w \boldsymbol{F}_{d,2})],
\end{equation}
where $\boldsymbol{i} =[1\ 0 \ 0]^T$, and $^1 \boldsymbol{R}_w$ denotes the inverse matrix of $^w\boldsymbol{R}_1$.

Finally, the nonlinear optimization model (\ref{eq:minPowerVarNoLimit}) is rewritten as follows:
\begin{equation}
    \label{eq:minPowerVar}
    \begin{aligned}
        &\min \limits_{\kappa_{1}} var (P_m(t)) \| _0 ^t\\
        &s.t.\ \kappa_{1}^{min}(f,\kappa_{2}) \leq \kappa_{1} \leq \kappa_{1}^{max}
    \end{aligned}.
\end{equation}

\section{Numerical simulation}
\label{sec:Simulation}
To validate the proposed dynamic model and optimization model, extensive numerical simulations are conducted in this section. Note that the parameters of simulation, including the physical parameters and hydrodynamic parameters of robotic fish, can be found in previous work \cite{Liao_2023_Tmech}. The physical parameters of the spring steel and transmission mechanism are presented in Table \ref{tab:others_Parameter}.


\begin{table*}[t]
\caption{The physical parameters of the spring steel and transmission mechanism} 
\begin{center}
\renewcommand{\arraystretch}{1.6}
\begin{tabular}{@{}*{9}{l}}
\toprule[0.5pt]
Name & $l_{T,1}$ & $l_{T,2}$  & $w_{T,1}$ & $w_{T,2}$ & $E_{i}$ & $J_{R}$ & $J_{S}$ & $J_{P}$ \\
\midrule[1pt]
Value & 0.083  &  0.020  &  0.028   & 0.025 & 197000 & $5.49\times 10^{-6}$  & $2.03\times 10^{-6}$ & $4.61\times 10^{-5}$ \\
Unit & m  &  m  &  m & m & MPa & $\rm kg\cdot m^2$  & $\rm kg\cdot m^2$  & $\rm kg\cdot m^2$\\
\bottomrule[0.5pt]
\end{tabular}
\end{center}
\label{tab:others_Parameter}
\end{table*}


\subsection{Simulation for fishtail swing}

\begin{figure}[t]
    \centering
    \includegraphics[width=8.8cm]{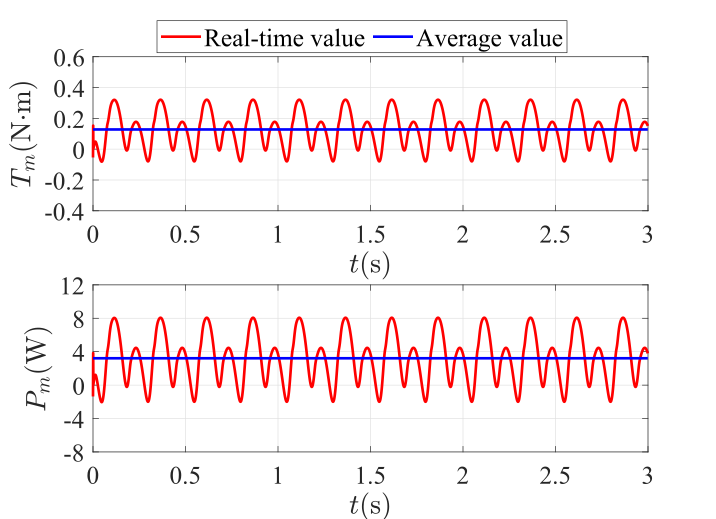}
    \caption{The real-time curves of motor torque $T_m$ and power $P_m$. The $f$, $\kappa_{1}$, and $\kappa_{2}$ are 4 Hz, 0.15 N$\cdot$m, and 1.31 N$\cdot$m, respectively.}
    \label{fig:RealTorqueAndPower}
\end{figure}

\begin{figure}[t]
    \centering
    \includegraphics[width=8.8cm]{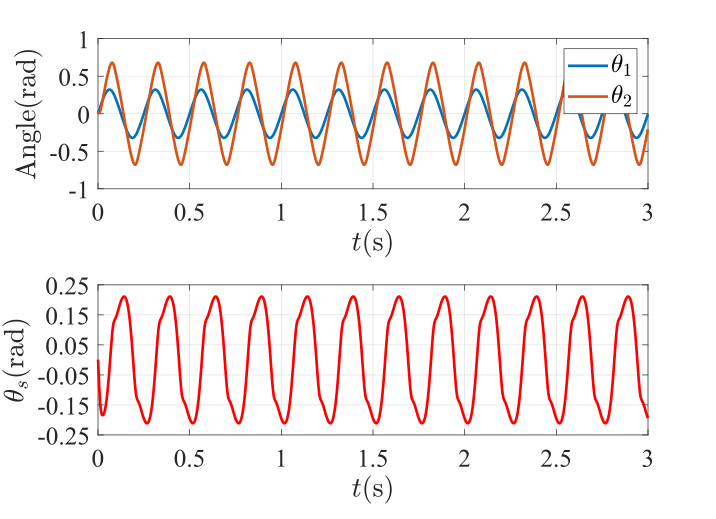}
    \caption{The curves of $\theta_1$, $\theta_2$ and $\theta_s$. The $f$, $\kappa_{1}$, and $\kappa_{2}$ are 4 Hz, 0.15 N$\cdot$m, and 1.31 N$\cdot$m, respectively. The three curves follow the sine-like curve, which is in line with the swing rule of the fishtail.}
    \label{fig:RealAngle}
\end{figure}

\begin{figure}[t]
    \centering
    \includegraphics[width=8.2cm]{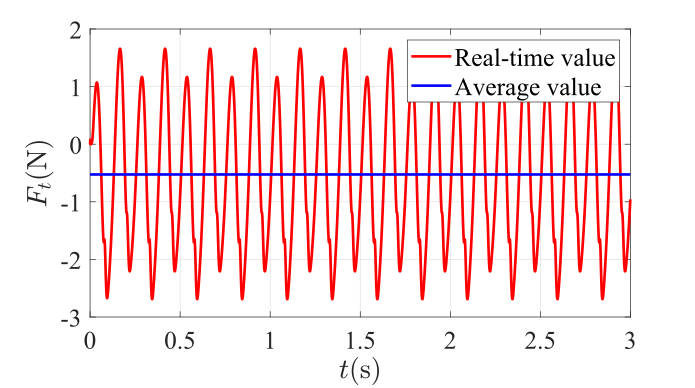}
    \caption{The curves of thrust for the fishtail. The $f$, $\kappa_{1}$, and $\kappa_{2}$ are 4 Hz, 0.15 N$\cdot$m, and 1.31 N$\cdot$m, respectively.}
    \label{fig:thrust}
\end{figure}

Based on the above dynamic model, the motor torque $T_m$ and power $P_m$ are obtained, as shown in Fig. \ref{fig:RealTorqueAndPower}, where $f$ = 4 Hz, $\kappa_{1}$ = 0.15 N$\cdot$m, and $\kappa_{2}$ = 1.31 N$\cdot$m. Due to the constant angular velocity of the motor, the curve of $T_m$ features the same law as that of $P_m$, and the average torque and the average power of the motor are 0.1277 N$\cdot$m and 3.21 W, respectively. The positive work of the motor is greater than the negative work, and the total work of the motor applied to drive the fishtail is greater than 0, which is in line with the objective law. Besides, according to the Fig. \ref{fig:RealAngle}, it's observed that the phase of $\theta_2$ lags behind that of $\theta_1$ due to the PES's elastic deformation, and the amplitude of $\theta_s$, $\theta_1$, and $\theta_2$ are 0.21 rad, 0.32 rad, and 0.68 rad, respectively. 

Figure \ref{fig:thrust} presents the thrust curve of robotic fish. The average thrust is -0.525 N, where the negative sign indicates the negative direction of the X-axis. Since the fishtail swings symmetrically, the oscillation frequency of the thrust is two times as large as the swing frequency of the fishtail. That is to say, when the fishtail finishes a drive stroke and recovery stroke, the thrust undergoes an oscillation cycle. In addition, as can be seen from Fig. \ref{fig:thrust}, the thrust of the fishtail is mainly produced during the recovery stroke, while the drive stroke mainly provides the drag force, which is in line with Fig. \ref{fig:AnalysisInoneCycle}. 

\begin{figure}[t]
    \centering
    \includegraphics[width=8.5cm]{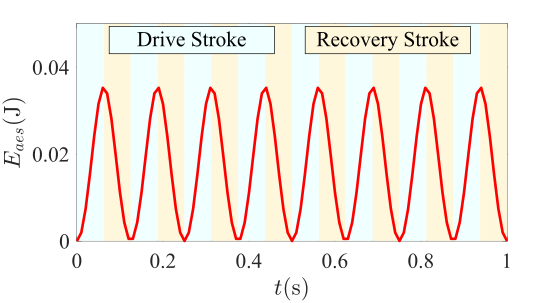}
    \caption{The stored energy of the AES versus time. The $f$, $\kappa_{1}$, and $\kappa_{2}$ are 4 Hz, 0.15 N$\cdot$m, and 1.31 N$\cdot$m, respectively.}
    \label{fig:RealEnergyOfAES}
\end{figure}

The stored energy $E_{aes}$ of the AES is depicted in Fig. \ref{fig:RealEnergyOfAES}, from which we can find that the period of the AES's energy-storing is 0.125 s, and the curve of $E_{aes}$ is similar to $\lvert \sin (8 \pi t) \lvert$. Obviously, the AES is capable of storing energy from the motor during the drive stroke, and the stored energy is released to drive the fishtail during the recovery stroke. The maximum elastic potential energy that the AES can store is about 0.035 J in this case.

\begin{figure}[t]
    \centering
    \includegraphics[width=8.8cm]{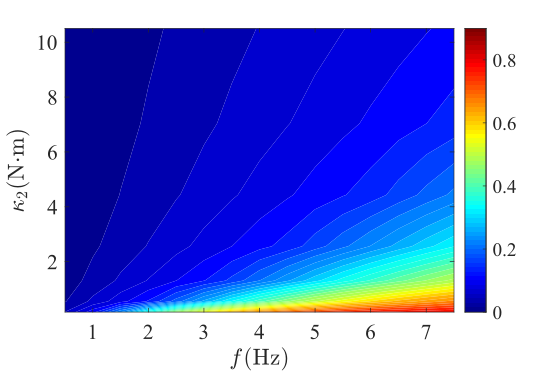}
    \caption{The PES's angle $\theta_s$ (rad) versus $f$ and $\kappa_{2}$.}
    \label{fig:AnglePFJoint}
\end{figure}

\begin{figure}[t]
    \centering
    \includegraphics[width=8.8cm]{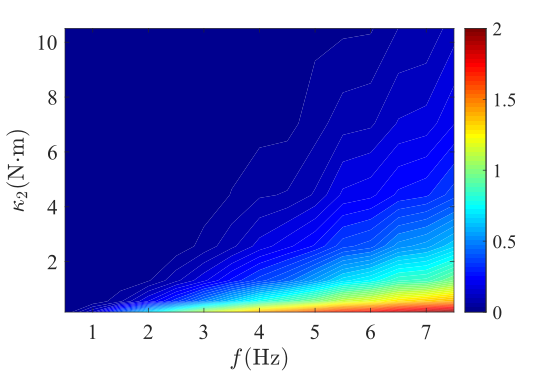}
    \caption{The phase difference $\phi_s$ (rad) between $\theta_1$ and $\theta_2$ versus $f$ and $\kappa_{2}$.}
    \label{fig:PhasePFJoint}
\end{figure}

To further validate the dynamic model, the changes of $\theta_s$, the phase difference $\phi_s$ between $\theta_1$ and $\theta_2$ are explored, as shown in Fig. \ref{fig:AnglePFJoint} and \ref{fig:PhasePFJoint}, respectively. When the $\kappa_{2}$ is relatively small (0.16 - 2.57 N$\cdot$m), $\theta_s$ and $\phi_s$ increase obviously with the increase of frequency, which is attributed to the increase of the PES's deformation magnitude due to the hydrodynamic forces of the caudal fin. For example, when the $\kappa_{2}$ is 0.16 N$\cdot$m, the minimum $\theta_s$ and $\phi_s$ are 0.059 rad and 0.003 rad, respectively, and the maximum $\theta_s$ and $\phi_s$ are 0.80 rad and 1.99 rad, respectively. The increments of the $\theta_s$ and $\phi_s$ are 0.741 rad and 1.987 rad, respectively. However, when $\kappa_{2}$ is relatively large (2.57 - 10.51 N$\cdot$m), $\theta_s$ increases slightly with the increases of frequency. At the $\kappa_{2}$ of 10.51 N$\cdot$m, $\theta_s$ increases from 0.007 rad to 0.12 rad, and the absolute increment is equal to 0.113 rad. Similarly, $\phi_s$ increases from 0.003 rad to 0.14 rad, and the absolute increment is equal to 0.137 rad. This phenomenon is because the PES's stiffness is very large, and it is difficult for the hydrodynamic force to make the PES bend and deform. It's obvious that the above results are consistent with the objective law, laying a solid foundation for the following optimization.

\subsection{Validation of AES’s stiffness range}

\begin{figure}[t]
    \centering
    \includegraphics[width=8.8cm]{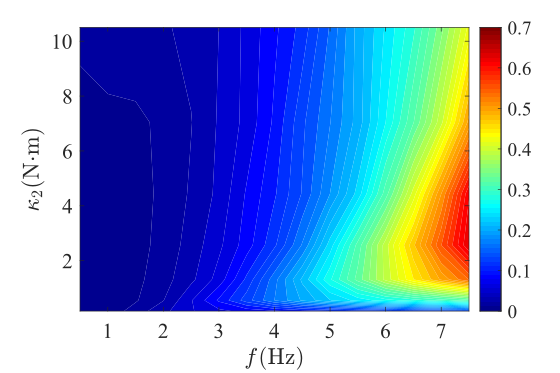}
    \caption{The curve of minimum $\kappa_{1}$ (N$\cdot$m) at the given $f$ and $\kappa_{2}$.}
    \label{fig:dT1min}
\end{figure}

In this section, we determine the maximum and minimum value of the $\kappa_{1}$. Firstly, according to the Eq. (\ref{eq:staticsTorque}), it's concluded that the bending moment for the AES increases due to the increase of the $\kappa_{1}$, resulting in the increase of the $\max(T_{m}^s)$. According to the Eq. (\ref{eq:maxdT1}), we can obtain the $\kappa_{1}^{max}$ of 14.46 N$\cdot$m according to the maximum output torque of the used motor of 3 N$\cdot$m. Secondly, for the different combinations of $f$ and $\kappa_{2}$, we determine the minimum $\kappa_{1}$ according to the Eq. (\ref{eq:mindT1}), as depicted in Fig. \ref{fig:dT1min}. There is a positive correlation between $\kappa_{1}^{min}(f,\kappa_{2})$ and $f$ at the given $\kappa_{2}$, which is attributed to the increase of the magnitude for the axial force acting on the AES. The minimum and maximum $\kappa_{1}^{min}(f,\kappa_{2})$ are 0.0011 N$\cdot$m and 0.64 N$\cdot$m, respectively. 

According to the Eq. (\ref{eq:AESEnergy}), we can conclude that the maximum $E_{aes}$ features a linear relationship with $\kappa_{1}$, and the AES's maximum energy for the minimum and maximum $\kappa_{1}$ are $2.5 \times 10^{-4}$ J and 3 J, respectively. Obviously, by adjusting the AES's stiffness, we can obtain a significant improvement in the energy-storing capability of the AES, which is why we adopt spring steel. The advantage is that when adjusting the AES's stiffness extensively to meet the actual requirements, we do not need to redesign the robotic fish to meet the assembly requirements, and the size and quality of the robotic fish also remain constant basically.

\subsection{Validation of AES’s stiffness optimization}

\begin{figure}[t]
\subfigure[$f$ = 4 Hz, $\kappa_{2}$ = 2.57 N$\cdot$m]{
\centering
\includegraphics[width=8.3cm]{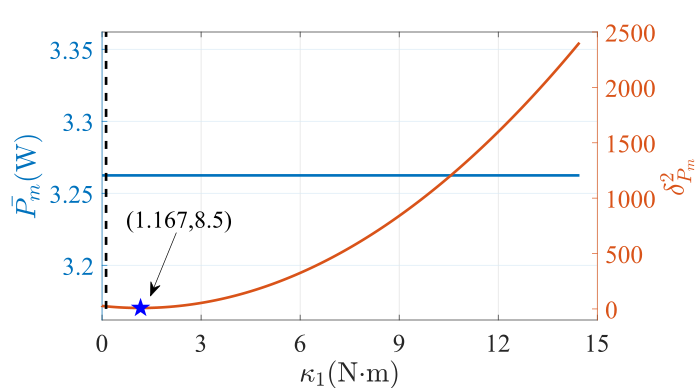}
\label{fig:AvePowerAndVar_4Hz_05mm}
}%
\centering

\subfigure[$f$ = 6 Hz, $\kappa_{2}$ = 10.51 N$\cdot$m]{
\centering
\includegraphics[width=8.3cm]{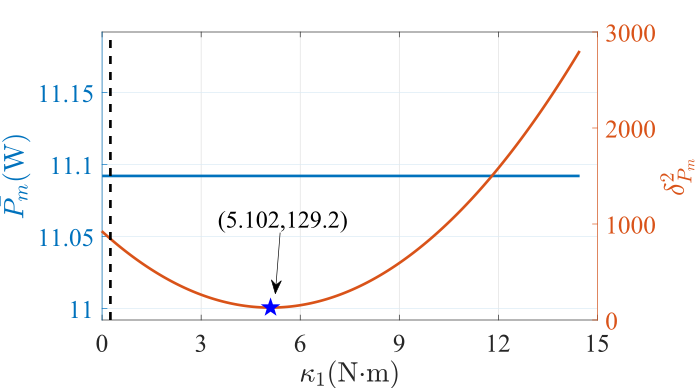}
\label{fig:AvePowerAndVar_6Hz_08mm}
}%
\centering
\caption{The average value $\bar P_m$ and variance $\delta^2_{P_m}$ of motor power versus $\kappa_{1}$. The black dotted line and the blue pentagram denote the $\kappa_{1}^{min}(f,\kappa_{2})$ and the optimal solution, respectively.}
\label{fig:AvePowerAndVar}
\end{figure}

Further, we explore the effect of $\kappa_{1}$ on the motor power at the given $f$ and $\kappa_{2}$, as shown in Fig. \ref{fig:AvePowerAndVar}. Obviously, $\kappa_{1}$ does not affect the average value $\bar P_m$ of the motor power, which validates that the AES’s stiffness does not affect the swimming speed of the robotic fish. The average torque of the motor is also independent of the $\kappa_{1}$ due to the constant angular velocity. However, $\kappa_{1}$ has a significant influence on the smoothness of the motor power, which is represented by the variance $\delta^2_{P_m}$. For example, according to Fig. \ref{fig:AvePowerAndVar_4Hz_05mm}, $\delta^2_{P_m}$ increases sharply at the given $f$ of 4 Hz and $\kappa_{2}$ of 2.57 N$\cdot$m, when $\kappa_{1}$ is greater than 3 N$\cdot$m. This is because the AES's bending moment increases sharply with the increase of the $\kappa_{1}$, which magnifies the peak value of the motor power and torque. 

From Fig. \ref{fig:AvePowerAndVar}, the optimal $\kappa_{1}$ and the peak value of $\tau_{J1}$ are 1.167 N$\cdot$m and 1.12 N$\cdot$m, respectively, when the $f$ and $\kappa_{2}$ are 4 Hz and 2.57 N$\cdot$m respectively. However, the optimal $\kappa_{1}$ and the peak value of $\tau_{J1}$ are 5.102 N$\cdot$m and 4.04 N$\cdot$m, respectively, at the given $f$ of 6 Hz and $\kappa_{2}$ of 10.51 N$\cdot$m. This phenomenon means that the peak value of $\tau_{J1}$ also increases with the increases of $f$ and $\kappa_{2}$, and in order to minimize the $\delta^2_{P_m}$, it is necessary to increase the AES's stiffness and $T_{e,1}$ correspondingly, to respond the change of $\tau_{J1}$. Besides, as can be seen from Fig. \ref{fig:AvePowerAndVar}, when $f$ and $\kappa_{2}$ are given, $\delta^2_{P_m}$ decreases firstly, and then increases with the increase of $\kappa_{1}$, and there is an optimal solution $\kappa_{1}^*$ to minimize the $\delta^2_{P_m}$.

\begin{figure}[t]
\subfigure[$\kappa_{1}$ = 0.35 N$\cdot$m]{
\centering
\includegraphics[width=8.5cm]{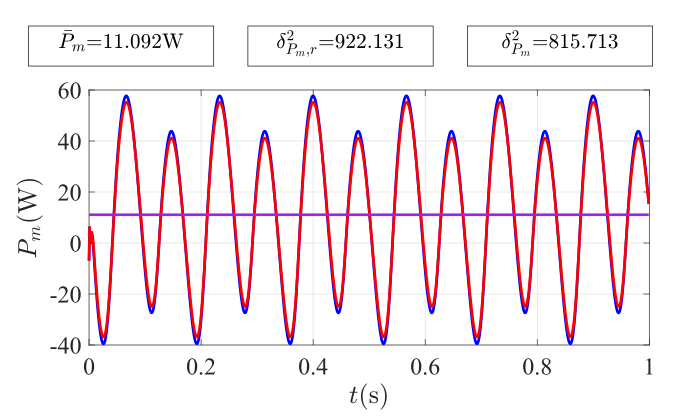}
\label{fig:realPower_6Hz_08mm_04mm}
}%
\centering

\subfigure[$\kappa_{1}$ = 5.102 N$\cdot$m]{
\centering
\includegraphics[width=8.5cm]{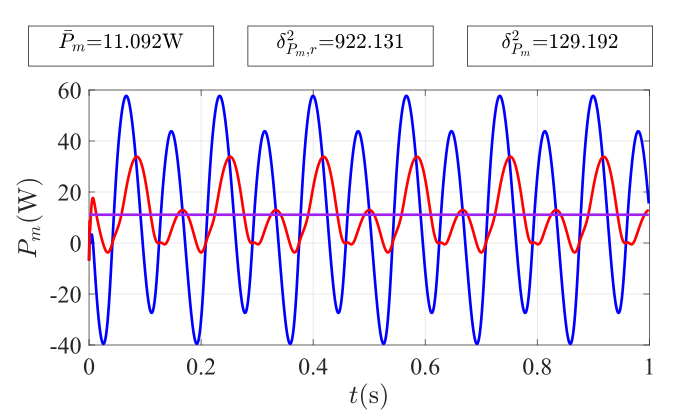}
\label{fig:realPower_6Hz_08mm_0973mm}
}%
\centering

\subfigure[$\kappa_{1}$ = 9.57 N$\cdot$m]{
\centering
\includegraphics[width=8.5cm]{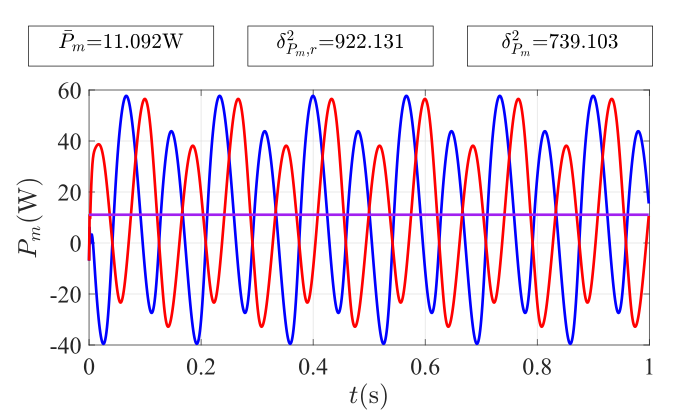}
\label{fig:realPower_6Hz_08mm_12mm}
}%
\centering
\caption{The real-time curves of motor power at different $\kappa_{1}$ within 0 - 1 s. The $f$ and $\kappa_{2}$ are 6 Hz and 10.51 N$\cdot$m, respectively. The red and blue curves represent the real-time motor power for the active-segment elastic spine with the given stiffness $\kappa_{1}$ and the active-segment rigid spine, respectively. Note that three blue curves are identical in three subgraphs. The purple line denotes the average power.}
\label{fig:realPower_6Hz_08mm_differentdT1}
\end{figure}

To intuitively present the influence of $\kappa_{1}$ on the motor power, Fig. \ref{fig:realPower_6Hz_08mm_04mm}, \ref{fig:realPower_6Hz_08mm_0973mm} and \ref{fig:realPower_6Hz_08mm_12mm} depict the real-time curves of the motor power when $\kappa_{1}$ are 0.35 N$\cdot$m, 5.102 N$\cdot$m, and 9.57 N$\cdot$m, respectively, where the $f$ and $\kappa_{2}$ are 6 Hz and 10.51 N$\cdot$m respectively. Obviously, the average power $\bar P_m$ of the motor is 11.092 W, which is independent of $\kappa_{1}$. By contrast, $\delta^2_{P_m}$ features great differences, and the minimum $\delta^2_{P_m}$ of 129.192 is obtained when $\kappa_{1}$ is 5.102 N$\cdot$m. Comparing Fig. \ref{fig:realPower_6Hz_08mm_04mm}, \ref{fig:realPower_6Hz_08mm_0973mm} and \ref{fig:realPower_6Hz_08mm_12mm}, it's found that when the variance of the motor power is small, the negative power is also significantly abated. In order to maintain the same average power, the peak value of the motor power is correspondingly weakened, thereby achieving the reduction of the $\delta^2_{P_m}$ and the stability improvement of the motor power. In addition, compared with the active-segment rigid spine that is incapable of storing energy, the AES can dent the peak value and fill the valley value for motor power by energy storage and release to respond to the fluctuations of the motor power effectively, if the $\kappa_{1}$ is adjusted reasonably, as shown in Fig. \ref{fig:realPower_6Hz_08mm_0973mm}. We use $\delta^2_{P_m,r}$ to denote the variance of the motor power for the active-segment rigid spine. Compared with the active-segment rigid spine, the $\delta^2_{P_m}$ for the AES with a stiffness of 5.102 N$\cdot$m decreases by 85.99\%. The appropriate $\kappa_{1}$ of the AES is necessary for the improvement of the power fluctuations. As can be seen from Fig. \ref{fig:realPower_6Hz_08mm_04mm} and \ref{fig:realPower_6Hz_08mm_12mm}, if $\kappa_{1}$ is relative small or great, $\delta^2_{P_m}$ is approximately equal to $\delta^2_{P_m,r}$, and the motor power still fluctuates violently. 

\begin{figure}[t]
    \centering
    \includegraphics[width=8.5cm]{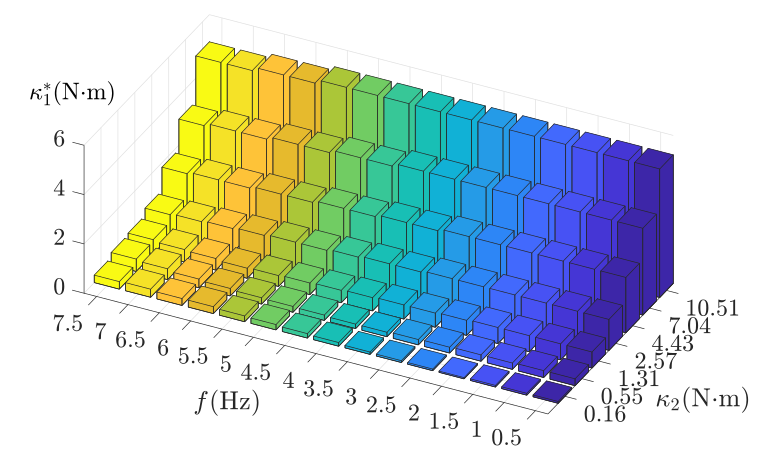}
    \caption{The optimal AES's stiffness $\kappa_{1}^*$ at different $f$ and $\kappa_{2}$.}
    \label{fig:OptimaldT1_dT2_f}
\end{figure}

\begin{figure}[t]
    \centering
    \includegraphics[width=8.6cm]{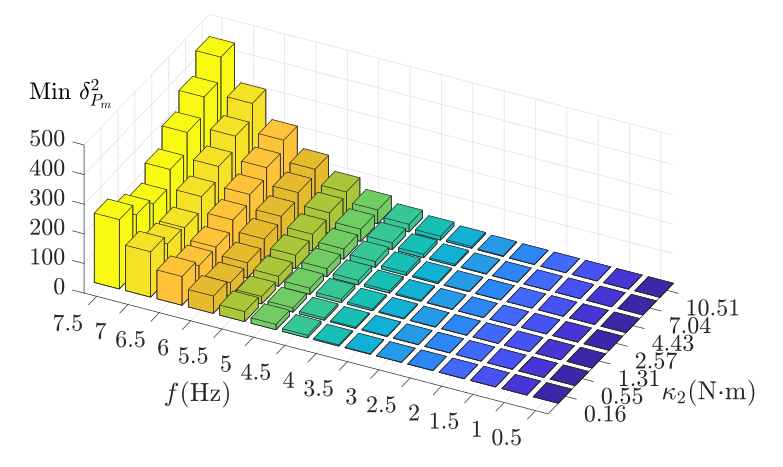}
    \caption{The minimum variance of motor power at different $f$ and $\kappa_{2}$.}
    \label{fig:OptimalVarPower_dT2_f}
\end{figure}

\begin{figure}[t]
    \centering
    \includegraphics[width=8.8cm]{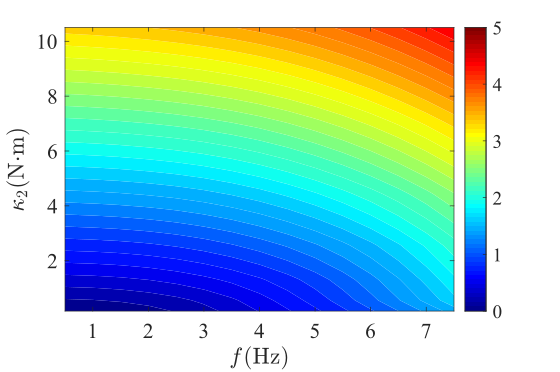}
    \caption{The peak value of $\tau_{J1}$ (N$\cdot$m) at different $f$ and $\kappa_{2}$.}
    \label{fig:MaxTaoJ1Versus_dT2_f}
\end{figure}

Further, $\kappa_{1}^*$ and the corresponding minimum variance are obtained by solving the Eq. (\ref{eq:minPowerVar}), as shown in Fig. \ref{fig:OptimaldT1_dT2_f} and \ref{fig:OptimalVarPower_dT2_f}, respectively. Figure \ref{fig:MaxTaoJ1Versus_dT2_f} depicts the peak value of $\tau_{J1}$ at different $f$ and $\kappa_{2}$. In Fig. \ref{fig:OptimaldT1_dT2_f}, \ref{fig:OptimalVarPower_dT2_f} and \ref{fig:MaxTaoJ1Versus_dT2_f}, we traverse all suitable values of $\kappa_{2}$ (corresponding to the thickness of 0.2 : 0.1 : 0.8 mm) to build a complete mapping relationship instead of focusing on the case at the optimal $\kappa_{2}$. According to Fig. \ref{fig:OptimaldT1_dT2_f} and \ref{fig:MaxTaoJ1Versus_dT2_f}, it's found that $\kappa_{1}^*$ increases with the increase of $f$ as the $\kappa_{2}$ is relatively small, which is because the AES's stiffness and bending moment need to be enhanced to respond the increase of the peak value of $\tau_{J1}$ for maintaining smaller variances. Besides, it can be seen that $\kappa_{1}^*$ increases at the given frequency as the $\kappa_{2}$ increases, which is also attributed to the increase of the peak value of $\tau_{J1}$. At the given ranges of $f$ and $\kappa_{2}$, the minimum and maximum values of $\kappa_{1}^*$ are 0.0443 N$\cdot$m and 5.21 N$\cdot$m, respectively, which are consistent with the selected stiffness range in practice basically.

\begin{figure}[t]
\subfigure[$\kappa_{2}$ = 1.31 N$\cdot$m]{
\centering
\includegraphics[width=8.5cm]{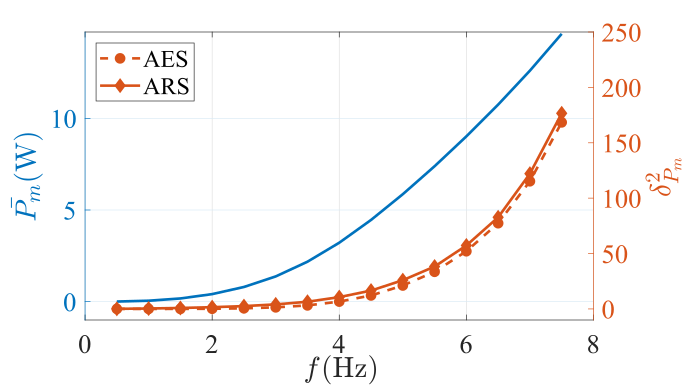}
\label{fig:ComparisionARSAndAES_04mm}
}%
\centering

\subfigure[$\kappa_{2}$ = 10.51 N$\cdot$m]{
\centering
\includegraphics[width=8.5cm]{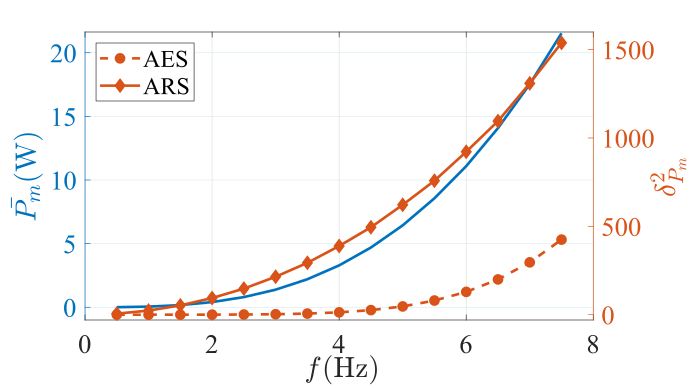}
\label{fig:ComparisionARSAndAES_08mm}
}%
\centering
\caption{The comparison of average value $\bar P_m$ and variance $\delta^2_{P_m}$ of motor power between the AES and ARS.} 
\label{fig:ComparisionARSAndAES}
\end{figure}

To validate the advantages of the AES's energy-storing, the comparisons between the AES and ARS in terms of the average value and variance of the motor power are conducted, as shown in Fig. \ref{fig:ComparisionARSAndAES}, where the variances for the AES are obtained by the nonlinear optimal model (\ref{eq:minPowerVar}). It's obvious that the average power of the motor for the AES is equal to that of the ARS, that is, the AES's energy-storing has no contribution to the propulsion performance of robotic fish. As can be seen from Fig. \ref{fig:ComparisionARSAndAES_04mm}, both of $\delta^2_{P_m}$ for the AES and ARS increase at the small $\kappa_{2}$ as the frequency increases, and the $\delta^2_{P_m}$ for AES is slightly smaller than that of the ARS. According to Fig. \ref{fig:MaxTaoJ1Versus_dT2_f}, it's found that the peak values of $\tau_{J1}$ are relatively small, and the maximum value is about 2 N$\cdot$m at the $\kappa_{2}$ of 1.31 N$\cdot$m, which results in the relatively small $\delta^2_{P_m}$ (less than 200). Hence, the AES makes a little contribution to reducing the $\delta^2_{P_m}$ by the advantages of energy-storing. Since $\delta^2_{P_m}$ is relatively large at the $\kappa_{2}$ of 10.51 N$\cdot$m owing to the large peak values of $\tau_{J1}$, the AES's energy-storing can adjust $\delta^2_{P_m}$ effectively.

\begin{figure}[t]
\subfigure[]{
\centering
\includegraphics[width=8.5cm]{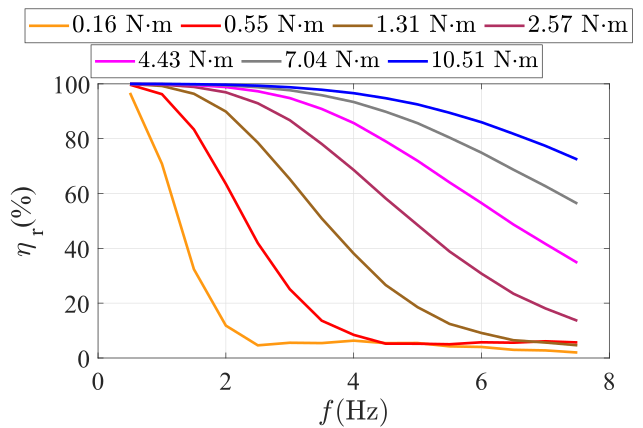}
\label{fig:RelativeDeclineRatio}
}%
\centering

\subfigure[]{
\centering
\includegraphics[width=8.5cm]{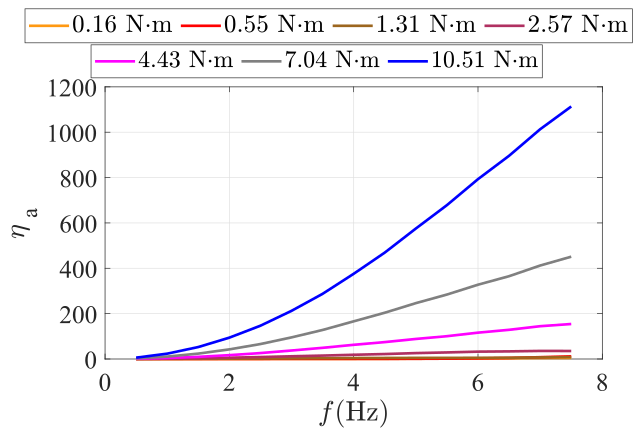}
\label{fig:AbsoluteDeclineRatio}
}%
\centering
\caption{The (a) relative decline ratio and (b) absolute decline value of the variance $\delta^2_{P_m}$ of motor power.} 
\label{fig:DeclineComparisionAESAndARS}
\end{figure}

To further evaluate the advantages of the AES's energy-storing, two indicators, i.e., the relative decline ratio $\eta_r$ and absolute decline value $\eta_a$ of the $\delta^2_{P_m}$ are defined as follows:
\begin{equation}
\eta_r = \frac{\delta^2_{P_m,r} - \delta^2_{P_m} }{\delta^2_{P_m,r}} \times 100\% ,
\end{equation}
\begin{equation}
\eta_a = \delta^2_{P_m,r} - \delta^2_{P_m}.
\end{equation}
For example, when the $f$ and $\kappa_{2}$ are 7.5 Hz and 10.51 N$\cdot$m, respectively, $\delta^2_{P_m,r}$ and $\delta^2_{P_m}$ are equal to 1538 and 424.9, respectively, and the $\eta_r$ and $\eta_a$ are 72.37\%, and 1113.1, respectively. Figures \ref{fig:RelativeDeclineRatio} and \ref{fig:AbsoluteDeclineRatio} depict the curves of $\eta_r$ and $\eta_a$ at the different combinations of $f$ and $\kappa_{2}$, respectively. It's obvious that, at the given frequency, the larger the $\kappa_{2}$ is, the larger the $\eta_r$ and $\eta_a$ are, which means that the AES's energy-storing plays a vital role in reducing the variance of the motor power. In addition, $\eta_r$ decreases and $\eta_a$ increases with the increase of frequency when the $\kappa_{2}$ is constant. Particularly, the relative decline ratio approaches 100\% as the frequency is relatively small.

\begin{figure}[t]
    \centering
    \includegraphics[width=8.2cm]{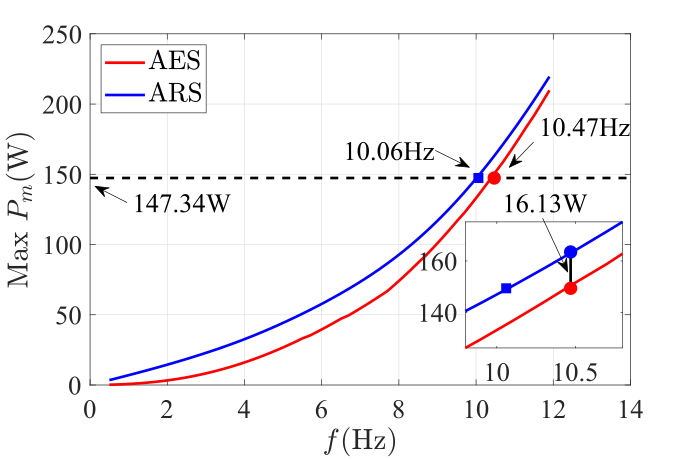}
    \caption{The comparison of the maximum power of motor between the AES and active-segment rigid spine. The $\kappa_{2}$ is 10.51 N$\cdot$m.}
    \label{fig:MaxPower_AES_ARS}
\end{figure}

\begin{figure}[t]
    \centering
    \includegraphics[width=8.6cm]{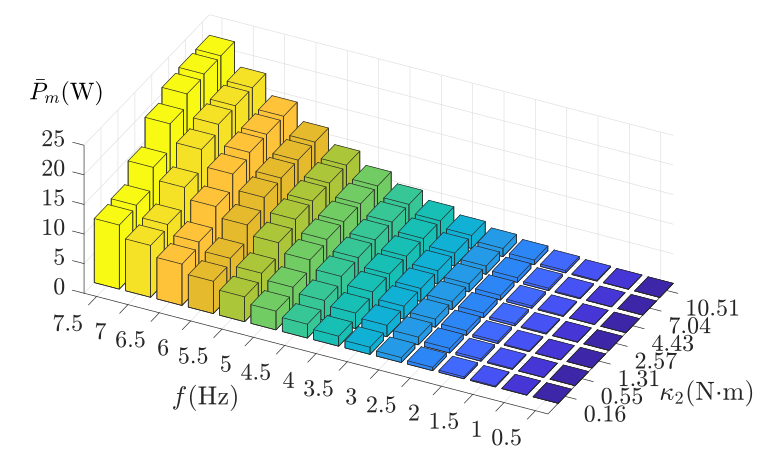}
    \caption{The average motor power $\bar P_m$ at different $f$ and $\kappa_{2}$.}
    \label{fig:AverPowerVersus_dT2_f}
\end{figure}

As analyzed above, the stable output power of the motor is beneficial to the improvement of the maximum frequency. Figure \ref{fig:MaxPower_AES_ARS} presents the maximum power of the motor for the AES and ARS, where the $\kappa_{2}$ is 10.51 N$\cdot$m, and $\kappa_{1}$ is equal to $\kappa_{1}^*$. Due to the allowable power value of the used motor of about 147.34 W, the highest frequencies of the motor for AES and ARS are about 10.47 Hz and 10.06 Hz, respectively, without damaging the motor. The increase ratio of frequency is 4.1\%, which is beneficial to the improvement of the propulsion performance. For example, the improvement of the thrust for the fishtail is from 3.60 N to 3.66 N, increasing by 0.06 N. Besides, when the frequency is 10.47 Hz, the maximum motor power decreases by 16.13 W (equivalent to 9.9\%) due to the AES’s energy-storing.

Finally, the average power of the motor at the different combinations of $f$ and $\kappa_{2}$ are explored, as shown in Fig. \ref{fig:AverPowerVersus_dT2_f}. $\bar P_m$ increases as the $f$ and $\kappa_{2}$ increase, and the maximum $\bar P_m$ of 21.53 W are obtained when the $f$ and $\kappa_{2}$ are 7.5 Hz and 10.51 N$\cdot$m, respectively. It's obvious that $\bar P_m$ will approach the allowable power value of the motor gradually if the $f$ and $\kappa_{2}$ increase further, which means that reducing $\delta^2_{P_m}$ and improving the stability of the motor power are very vital for protecting the motor from being damaged. 

\subsection{Discussion}

Based on the above analyses, we can conclude that the AES's energy-storing is incapable of improving the swimming performances of the robotic fish in terms of swimming speed and propulsion efficiency, but is beneficial to improving the stability of the motor power by adjusting the AES's stiffness reasonably. Moreover, there is an optimal AES's stiffness that can minimize the variance of the motor power at the given frequency and PES's stiffness, and the optimal AES's stiffness is positively correlated with the peak value of the equivalent joint torque.

Besides, compared with the ARS, the advantages of AES's energy-storing can be fully reflected in the improvement of the smoothness of motor power. On the one hand, it's beneficial to protect the motor from damage and improve the motor's service life. On the other hand, the motor's performance in terms of maximum frequency can be improved effectively. Concretely, in the case of using the same motor, since the ARS cannot store energy, the peak power of the motor exceeds the allowable value easily, which limits the maximum average power and frequency. Benefiting from the AES's energy-storing, the motor can achieve higher frequency and average power to improve the swimming performance of robotic fish without the damage to motor by adjusting the AES's stiffness reasonably.

\section{Conclusions and future works}
\label{sec:Conclusions}
In this paper, we propose a wire-driven elastic robotic fish, whose fishlike tail is based on dual spring steel and includes an AES and a PES. As the fishtail swings, the AES can generate bending deformation under the action of the wire driving and motor, to realize periodic energy storage. We analyze and verify that the AES's energy-storing and stiffness have no contribution to the swimming performance of the robotic fish, but can affect the smoothness of the motor power. Hence, with the aim of minimizing the variance of the output power of the motor, we optimize the AES's stiffness by the developed nonlinear optimization model that is based on the Lagrangian dynamics and cantilever beam model. The results show that, compared with the ARS, the optimized AES can effectively smooth the motor power, and reduce the peak value as well as variance of the motor power. Hence, the motor power does not exceed the allowable power value easily, which is beneficial to protect the motor from damage and improve the service life of the motor.

In the future, we will focus on how to optimize the stiffness of the passive-segment elastic spine and active-segment elastic spine synchronously, and how to design a flexible fishtail that can adjust the stiffness of the AES and PES online.

\section*{Acknowledgments}

This work was supported by the National Nature Science Foundation of China (Grant numbers: 62033013, 62003341, 62003342, 62203436).

\section*{Conflict of interests} 

The authors have no relevant financial or non-financial interests to disclose.

\section*{Data availability statement}

All data and codes for the current study are available from the corresponding author on reasonable request.

\section*{Author contributions}

Chao Zhou designed this study. Xiaocun Liao implemented the algorithms and simulations. All authors, including Xiaocun Liao, Chao Zhou, Junfeng Fan, Zhuoliang Zhang, Zhaoran Yin, and Liangwei Deng, contributed to the writing of the manuscript, and approved the final manuscript.

\vspace{2em}
\textbf{Publisher’s Note}\ \ \ Springer Nature remains neutral with regard
to jurisdictional claims in published maps and institutional affiliations.

\vspace{1em}
\textbf{Xiaocun Liao}\ \ \  received the B.E. degree in Detection, Guidance and Control Technology from Central South University (CSU), Changsha, China, in 2019. He is currently working toward the Ph.D. degree in control theory and control engineering with the Institute of Automation, Chinese Academy of Sciences (IACAS), Beijing, China. His research interests include the bioinspired robot fish and intelligent control systems.

\vspace{1em}
\textbf{Chao Zhou}\ \ \  received the B.E. degree in automation from Southeast University, Nanjing, China, in July 2003, and the Ph.D. degree in control theory and control engineering from the Institute of Automation, Chinese Academy of Sciences (IACAS), Beijing, China, in 2008. He is currently a Professor with the State Key Laboratory of Management and Control for Complex Systems, IACAS. His current research interests include the motion control of robot, the bioinspired robotic fish, and embedded system of robot.

\vspace{1em}
\textbf{Junfeng Fan}\ \ \  received the B.S. degree in mechanical engineering and automation from Beijing Institute of Technology, Beijing, China, in 2014 and Ph.D. degree in control theory and control engineering from the Institute of Automation, Chinese Academy of Sciences (IACAS), Beijing, China, in 2019. He is currently an Associate Professor of Control Theory and Control Engineering with the State Key Laboratory of Management and Control for Complex Systems, IACAS, Beijing. His research interests include robot vision and underwater robot.

\vspace{1em}
\textbf{Zhuoliang Zhang}\ \ \  received the B.E. degree in automation from Tongji University, Shanghai, China, in July 2018. He is currently pursuing the Ph.D. degree in control theory and control engineering with the Institute of Automation, Chinese Academy of Sciences, and also with the University of Chinese Academy of Sciences, Beijing. His research interests include measurements, sensor signal processing, and intelligent control.

\vspace{1em}
\textbf{Zhaoran Yin}\ \ \  received the B.E. degree in Automation from the School of Control Science and Engineering, Shandong University, Jinan, China, in 2017, and the M.S. degree in Electrical Engineering from the University of Texas at Dallas, Richardson, TX, USA. He is currently working toward the Ph.D. degree in control theory and control engineering with the Institute of Automation, Chinese Academy of Sciences (IACAS), Beijing, China. His research interests include bioinspired underwater robots and intelligent control systems.

\vspace{1em}
\textbf{Liangwei Deng}\ \ \   received the B.E. degree in mechanical design manufacture and automation from University Of Electronic Science And Technology Of China (UESTC), Chengdu, China, in 2020. He is currently working toward the Ph.D. degree in control theory and control engineering with the Institute of Automation, Chinese Academy of Sciences (IACAS), Beijing, China. His research interests include bioinspired robots.

\end{CJK}

\begin{thebibliography}{99}
\balance
\bibitem{Scaradozzi_2017_OE} Scaradozzi, D., Palmieri, G., Costa, D., Pinelli, A.: BCF swimming locomotion for autonomous underwater robots: a review and a novel solution to improve control and efficiency. Ocean Engineering. \textbf{130}, 437-453 (2017). \href{https://doi.org/10.1016/j.oceaneng.2016.11.055}{https://doi.org/10.1016/j.oceaneng.2016.11.055}

\bibitem{Zhang_2020_NODY} Zhang, P., Wu, Z., Meng, Y., Tan, M., Yu, J.: Nonlinear model predictive position control for a tail-actuated robotic fish. Nonlinear Dyn. \textbf{101}, 2235–2247 (2020). \href{https://doi.org/10.1007/s11071-020-05963-2}{https://doi.org/10.1007/s11071-020-05963-2}

\bibitem{Omari_2021_OE} Omari, M., Ghommem, M., Romdhane, L., Hajj, M.R.: Performance analysis of bio-inspired transformable robotic fish tail. Ocean Engineering. \textbf{244}, 110406 (2022). \href{https://doi.org/10.1016/j.oceaneng.2021.110406}{https://doi.org/10.1016/j.oceaneng.2021.110406}

\bibitem{Clapham_2015_Springer} Clapham, R.J., Hu, H.: iSplash: Realizing Fast Carangiform Swimming to Outperform a Real Fish. In: Du, R., Li, Z., Youcef-Toumi, K., and Valdivia y Alvarado, P. (eds.) Robot Fish. pp. 193–218. Springer Berlin Heidelberg, Berlin, Heidelberg (2015). \href{https://doi.org/10.1007/978-3-662-46870-8_7}{https://doi.org/10.1007/978-3-662-46870-8\_7}

\bibitem{Liu_2022_BB} Liu, S., Wang, Y., Li, Z., Jin, M., Ren, L., Liu, C.: A fluid-driven soft robotic fish inspired by fish muscle architecture. Bioinspir. Biomim. \textbf{17}(2), 026009 (2022). \href{https://doi.org/10.1088/1748-3190/ac4afb}{https://doi.org/10.1088/1748-3190/ac4afb}

\bibitem{Aubin_2019_Nature} Aubin, C.A., Choudhury, S., Jerch, R., Archer, L.A., Pikul, J.H., Shepherd, R.F.: Electrolytic vascular systems for energy-dense robots. Nature. \textbf{571}(7763), 51–57 (2019). \href{https://doi.org/10.1038/s41586-019-1313-1}{https://doi.org/10.1038/s41586-019-1313-1}

\bibitem{Marchese_2014_SoftRobotics}  Marchese, A.D., Onal, C.D., Rus, D.: Autonomous Soft Robotic Fish Capable of Escape Maneuvers Using Fluidic Elastomer Actuators. Soft Robotics. \textbf{1}(1), 75–87 (2014). \href{https://doi.org/10.1089/soro.2013.0009}{https://doi.org/10.1089/soro.2013.0009}

\bibitem{Katzschmann_2018_SciRob} Katzschmann, R.K., DelPreto, J., MacCurdy, R., Rus, D.: Exploration of underwater life with an acoustically controlled soft robotic fish. Sci. Robot. \textbf{3}(16), eaar3449 (2018). \href{https://doi.org/10.1126/scirobotics.aar3449}{https://doi.org/10.1126/scirobotics.aar3449}

\bibitem{Wang_2008_SMS} Wang, Z., Hang, G., Wang, Y., Li, J., Du, W.: Embedded SMA wire actuated biomimetic fin: a module for biomimetic underwater propulsion. Smart Mater. Struct. \textbf{17}(2), 025039 (2008). \href{https://doi.org/10.1088/0964-1726/17/2/025039}{https://doi.org/10.1088/0964-1726/17/2/025039}

\bibitem{Li_2022_SMS} Li, L., Guo, X., Liu, Y., Zhang, D., Liao, W.-H.: Dynamic modeling of a fish tail actuated by IPMC actuator based on the absolute nodal coordinate formulation. Smart Mater. Struct. \textbf{31}(11), 115005 (2022). \href{https://doi.org/10.1088/1361-665X/ac8c0a}{https://doi.org/10.1088/1361-665X/ac8c0a}


\bibitem{Li_2021_Nature} Li, G., Chen, X., Zhou, F., Liang, Y., Xiao, Y., Cao, X., Zhang, Z., Zhang, M., Wu, B., Yin, S., Xu, Y., Fan, H., Chen, Z., Song, W., Yang, W., Pan, B., Hou, J., Zou, W., He, S., Yang, X., Mao, G., Jia, Z., Zhou, H., Li, T., Qu, S., Xu, Z., Huang, Z., Luo, Y., Xie, T., Gu, J., Zhu, S., Yang, W.: Self-powered soft robot in the Mariana Trench. Nature. \textbf{591}(7848), 66–71 (2021). \href{https://doi.org/10.1038/s41586-020-03153-z}{https://doi.org/10.1038/s41586-020-03153-z}


\bibitem{Ning_2022_MRS} Ning, K., Hartono, P., Sawada, H.: Using inverse learning for controlling bionic robotic fish with SMA actuators. MRS Advances. \textbf{7}, 649–655 (2022). \href{https://doi.org/10.1557/s43580-022-00328-w}{https://doi.org/10.1557/s43580-022-00328-w}


\bibitem{Chen_2020_SMC} Chen, X., Yu, J., Wu, Z., Meng, Y., Kong, S.: Toward a Maneuverable Miniature Robotic Fish Equipped With a Novel Magnetic Actuator System. IEEE Trans. Syst. Man Cybern, Syst. \textbf{50}(7), 2327–2337 (2020). \href{https://doi.org/10.1109/TSMC.2018.2812903}{https://doi.org/10.1109/TSMC.2018.2812903}

\bibitem{Huang_2021_BB} Huang, C., Lai, Z., Zhang, L., Wu, X., Xu, T.: A magnetically controlled soft miniature robotic fish with a flexible skeleton inspired by zebrafish. Bioinspir. Biomim. \textbf{16}(6), 065004 (2021). \href{https://doi.org/10.1088/1748-3190/ac23a9}{https://doi.org/10.1088/1748-3190/ac23a9}


\bibitem{Chen_2019_SciRob} Chen, B., Jiang, H.: Swimming performance of a tensegrity robotic fish. Soft robotics. \textbf{6}(4), 520-531 (2019). \href{https://doi.org/10.1089/soro.2018.0079}{https://doi.org/10.1089/soro.2018.0079}

\bibitem{Chen_2021_TRO} Chen, B., Jiang, H.: Body Stiffness Variation of a Tensegrity Robotic Fish Using Antagonistic Stiffness in a Kinematically Singular Configuration. IEEE Trans. Robot. \textbf{37}(5), 1712–1727 (2021). \href{https://doi.org/10.1109/TRO.2021.3049430}{https://doi.org/10.1109/TRO.2021.3049430}

\bibitem{Zheng_2013_ICIRS} Li, Z., Zhong, Y., Du, R.: A novel underactuated wire-driven robot fish with vector propulsion. In: 2013 IEEE/RSJ International Conference on Intelligent Robots and Systems. pp. 941–946. IEEE, Tokyo (2013). \href{https://doi.org/10.1109/IROS.2013.6696463}{https://doi.org/10.1109/IROS.2013.6696463}

\bibitem{Zhong_2017_Tmech} Zhong, Y., Li, Z., Du, R.: A Novel Robot Fish With Wire-Driven Active Body and Compliant Tail. IEEE/ASME Trans. Mechatron. \textbf{22}(4), 1633–1643 (2017). \href{https://doi.org/10.1109/TMECH.2017.2712820}{https://doi.org/10.1109/TMECH.2017.2712820}

\bibitem{LiuJ_2021_BB} Liu, J., Zhang, C., Liu, Z., Zhao, R., An, D., Wei, Y., Wu, Z., Yu, J.: Design and analysis of a novel tendon-driven continuum robotic dolphin. Bioinspir. Biomim. \textbf{16}(6), 065002 (2021). \href{https://doi.org/10.1088/1748-3190/ac2126}{https://doi.org/10.1088/1748-3190/ac2126}

\bibitem{Shintake_2020_ICRA} Shintake, J., Zappetti, D., Peter, T., Ikemoto, Y., Floreano, D.: Bio-inspired Tensegrity Fish Robot. In: 2020 IEEE International Conference on Robotics and Automation (ICRA). pp. 2887–2892. IEEE, Paris, France (2020). \href{https://doi.org/10.1109/ICRA40945.2020.9196675}{https://doi.org/10.1109/ICRA40945.2020.9196675}

\bibitem{Estarki_2021_ICRoM}  Estarki, M., Varnousfaderani, R.H., Ghafarirad, H., Zareinejad, M.: Design and Implementation of a Soft Robotic Fish Based on Carangiform Fish Swimming. In: 2021 9th RSI International Conference on Robotics and Mechatronics (ICRoM). pp. 322–328. IEEE, Tehran, Iran, Islamic Republic of (2021). \href{https://doi.org/10.1109/ICRoM54204.2021.9663484}{https://doi.org/10.1109/ICRoM54204.2021.9663484}

\bibitem{Park_2010_ICBRB} Park, Y.-J., Jeong, U., Lee, J., Kim, H.-Y., Cho, K.-J.: The effect of compliant joint and caudal fin in thrust generation for robotic fish. In: 2010 3rd IEEE RAS \& EMBS International Conference on Biomedical Robotics and Biomechatronics. pp. 528–533. IEEE, Tokyo, Japan (2010). \href{https://doi.org/10.1109/BIOROB.2010.5626007}{https://doi.org/10.1109/BIOROB.2010.5626007}

\bibitem{Reddy_2018_MMT} Reddy N, S., Sen, S., Har, C.: Effect of flexural stiffness distribution of a fin on propulsion performance. Mechanism and Machine Theory. \textbf{129}, 218–231 (2018). \href{https://doi.org/10.1016/j.mechmachtheory.2018.07.012}{https://doi.org/10.1016/j.mechmachtheory.2018.07.012}

\bibitem{Chen_2022_Tmech} Chen, D., Wu, Z., Meng, Y., Tan, M., Yu, J.: Development of a High-Speed Swimming Robot With the Capability of Fish-Like Leaping. IEEE/ASME Trans. Mechatron. \textbf{27}(5), 3579-3589 (2022). \href{https://doi.org/10.1109/TMECH.2021.3136342}{https://doi.org/10.1109/TMECH.2021.3136342}


\bibitem{LiZ_2013_AMM} Li, Z., Du, R.X., Zhang, Y., Li, H.: Robot Fish with Novel Wire-Driven Continuum Flapping Propulsor. AMM. \textbf{300-301}, 510–514 (2013). \href{https://doi.org/10.4028/www.scientific.net/AMM.300-301.510}{https://doi.org/10.4028/www.scientific.net/AMM.300-301.510}

\bibitem{Lau_2015_RAM} Lau, W.P., Zhong, Y., Du, R., Li, Z.: Bladderless swaying wire-driven Robot Shark. In: 2015 IEEE 7th International Conference on Cybernetics and Intelligent Systems (CIS) and IEEE Conference on Robotics, Automation and Mechatronics (RAM). pp. 155–160. IEEE, Siem Reap, Cambodia (2015). \href{https://doi.org/10.1109/ICCIS.2015.7274613}{https://doi.org/10.1109/ICCIS.2015.7274613}

\bibitem{Fujiwara_2017_Abmech} Fujiwara, S., Yamaguchi, S.: Development of Fishlike Robot that Imitates Carangiform and Subcarangiform Swimming Motions. J. Abmech. \textbf{6}(1), 1–8 (2017). \href{https://doi.org/10.5226/jabmech.6.1}{https://doi.org/10.5226/jabmech.6.1}

\bibitem{ElDaou_2012_ICRA} El Daou, H., Salumae, T., Toming, G., Kruusmaa, M.: A bio-inspired compliant robotic fish: Design and experiments. In: 2012 IEEE International Conference on Robotics and Automation. pp. 5340–5345. IEEE, St Paul, MN, USA (2012). \href{https://doi.org/10.1109/ICRA.2012.6225321}{https://doi.org/10.1109/ICRA.2012.6225321}


\bibitem{Valdivia_2006_JDS} Valdivia y Alvarado, P., Youcef-Toumi, K.: Design of Machines With Compliant Bodies for Biomimetic Locomotion in Liquid Environments. Journal of Dynamic Systems, Measurement, and Control. \textbf{128}(1), 3–13 (2006). \href{https://doi.org/10.1115/1.2168476}{https://doi.org/10.1115/1.2168476}

\bibitem{Jiang_2019_CAC} Jiang, Y., Liu, X., Chen, H., Gong, W., Lu, Y., Zhang, W.: Design and Modeling of a Biomimetic Wire-driven Soft Robotic Fish. In: 2019 Chinese Automation Congress (CAC). pp. 1778–1782. IEEE, Hangzhou, China (2019). \href{https://doi.org/10.1109/CAC48633.2019.8996663}{https://doi.org/10.1109/CAC48633.2019.8996663}

\bibitem{Liao_2023_Tmech} Liao, X., Zhou, C., Wang, J., Tan. M.: A Wire-Driven Dual Elastic Fishtail With Energy Storing and Passive Flexibility IEEE/ASME Trans. Mechatron (2023). doi: \href{https://doi.org/10.1109/TMECH.2023.3318219}{https://doi.org/10.1109/TMECH.2023.3318219}

\end{thebibliography}
\end{document}